\documentclass[twocolumn]{aastex631}

\usepackage{graphicx,url,twoopt,natbib}
\usepackage{amsmath}
\usepackage{hyperref}                
\usepackage{xcolor}
\usepackage{multirow}

\submitjournal{AJ}

\begin{document}

\title{Serial MultiView: an efficient approach to mitigating atmospheric spatial-structure errors for VLBI astrometry}

\correspondingauthor{Bo Zhang}
\email{zb@shao.ac.cn}

\author[0000-0002-9768-2700]{Jingdong Zhang}
\affiliation{Shanghai Astronomical Observatory, Chinese Academy of Sciences, 80 Nandan Road, Shanghai 200030, People’s Republic of China}
\affiliation{University of Chinese Academy of Sciences, No.19 (A) Yuquan Rd, Shijingshan, Beijing 100049, People’s Republic of China}

\author[0000-0003-1353-9040]{Bo Zhang}
\affiliation{Shanghai Astronomical Observatory, Chinese Academy of Sciences, 80 Nandan Road, Shanghai 200030, People’s Republic of China}

\author[0000-0003-2953-6442]{Shuangjing Xu}
\affiliation{Korea Astronomy and Space Science Institute, 776 Daedeok-daero,
Yuseong-gu, Daejeon 34055, Republic of Korea}
\affiliation{Shanghai Astronomical Observatory, Chinese Academy of Sciences, 80 Nandan Road, Shanghai 200030, People’s Republic of China} 

\author[0000-0003-4871-9535]{Maria J. Rioja}
\affiliation{International Centre for Radio Astronomy Research, The University of Western Australia, 35 Stirling Hwy, Crawley, WA, Australia}
\affiliation{Observatorio Astronómico Nacional (IGN), Alfonso XII, 3 y 5, 28014, Madrid, Spain}

\author[0000-0003-0392-3604]{Richard Dodson}
\affiliation{International Centre for Radio Astronomy Research, The University of Western Australia, 35 Stirling Hwy, Crawley, WA, Australia}

\author[0000-0001-7573-0145]{Xiaofeng Mai}
\affiliation{Shanghai Astronomical Observatory, Chinese Academy of Sciences, 80 Nandan Road, Shanghai 200030, People’s Republic of China}
\affiliation{University of Chinese Academy of Sciences, No.19 (A) Yuquan Rd, Shijingshan, Beijing 100049, People’s Republic of China}
\affiliation{University of Helsinki, P.O. Box 64, FI-00014, Finland}

\author[0000-0003-1751-676X]{Oleg Titov}
\affiliation{Geoscience Australia, Canberra, ACT, 2601, Australia}



\begin{abstract}

Atmospheric propagation errors are a main constraint on the accuracy of Very Long Baseline Interferometry (VLBI) astrometry.
For relative astrometry, differential techniques can mitigate these errors, but their effectiveness diminishes with decreasing elevation and increasing angular separations between target and calibrator, among others.
The MultiView technique addresses atmospheric spatial-structure errors by observing multiple calibrators around the target and interpolating at the target position, thereby reducing atmospheric errors more effectively than phase-referencing with only one calibrator.
The first MultiView realisation at 1.6\,GHz involved cyclically observing all calibrators and the target, fitting a phase plane from calibrator solutions in each cycle, and is a well-established technique.
This implementation reduces on-target time and is constricted by the short atmospheric coherence time at high frequencies.
We propose a new realisation, serial MultiView, which rotates the phase plane iteratively based on the time series of calibrator residual phases.
This new strategy obviates the necessity of observing all calibrators within each cycle, thereby shortening the observing cycle and offering considerable potential at higher frequencies where the temporal structure is the dominant source of errors.
Additionally, by incorporating time-domain information in the iterations, phase ambiguities can be accurately and automatically identified.
We verify the astrometric accuracy of serial MultiView at 5\,GHz by comparing it to conventional MultiView, achieving $<$10\,$\mu$as error in RA direction, and show the calibration overhead can be reduced in both approaches.
This approach enables efficient, high-accuracy differential astrometry and artifact-reduced imaging for astrophysical studies, and we provide a user-friendly tool for it.

\end{abstract}



\section{Introduction}     \label{sec:introduction}

Very Long Baseline Interferometry (VLBI) provides the highest differential astrometric accuracy at present.
The key to this high accuracy is the calibration of the interferometric phase observable.
Phase-referencing \citep[PR,][]{1990AJ.....99.1663L, 1995ASPC...82..327B} techniques, in short observing the target and a nearby calibrator alternatively or simultaneously, are used for the calibration.
Most of the error terms are canceled out, benefiting from the small angular separation between target and calibrator.
However, the thermal noise limit of VLBI astrometry has only been reached with PR in a limited number of cases.
This is partly because of the residual delay caused by the atmosphere, especially when a close enough calibrator is not available.
The atmospheric residual delay includes tropospheric and ionospheric terms, which have a time-varying nature and irregular spatial structures.
At higher frequencies, tropospheric turbulence is the dominant term; while at lower frequencies, irregularities in ionospheric plasma density dominate.

The MultiView technique \citep{2017AJ....153..105R} compensates for atmospheric spatial-structure errors through observations of multiple calibrators around the target, and demonstrates a significant astrometric improvement at 1.6 GHz with the potential of wide applicability to many sources.
Basically, in MultiView, the atmospheric spatial structure is approximated to be a plane (in the simplest case), and the phase correction for the target is interpolated from the phases of three or more calibrators.
There are some early attempts that applied similar phase interpolation techniques, e.g., \citet{2002evn..conf...57R,2002evn..conf...53F,2003ApJ...598..704F,2006PASJ...58..777D}.

In recent years, this technique has become more popular with an increasing number of successful applications which approached thermal noise errors with existing instruments, and new implementations are being developed to adapt its scientific application to particular domains.
For example, inverse MultiView, for the cases when the targets are strong quasars \citep[8.3\,GHz]{2022ApJ...932...52H} or masers \citep[6.7\,GHz]{2023ApJ...953...21H};
\citet{2023MNRAS.519.4982D} managed to apply phase interpolation to pulsar VLBI astrometry.
Additionally, it is a very active field of research in the era of next-generation interferometers and is a main driver for key multi-pixel capabilities that enable simultaneous observations of the target and multiple calibrators \citep{2020AARv..28....6R}.

Conventional Multiview (cMV) introduced in \citet{2017AJ....153..105R} is optimized for precise astrometric studies in observations where the atmospheric spatial structure changes are dominant, e.g. low frequencies ($<$5\,GHz), low elevations and/or the longest baselines, among others.
The scheduling constraints for precise sampling of the dominant spatial structure results in a high density of calibration scans within the coherence time, i.e. it imposes a high calibration overhead leading to low fractional on-target time, in general.
At mid-frequencies ($\sim$5\,--\,8\,GHz), where the atmospheric temporal structure becomes significant, such a sequence can be surplus to requirements;
In these cases alternative schedules with greater on-target observing time could be more effective and have the potential for improved results.
The limit of applicability of this method is when the coherence time is shorter than the duration of the calibrator scans sequence, i.e., high frequencies ($>$8\,GHz),
Finally, the successful analysis using multiple calibrators requires phase connections between the calibrator scans used to define each plane.  A-priori errors can lead to incorrect phase connection, or ``phase ambiguity'' problems (i.e. wrapping past $\pm 2\pi$), due to the large residual delays, as discussed in \citet[][Sect.~3-4]{2022PASP..134l3001R}.

In this paper, we propose a new approach, serial MultiView (sMV), to expand the benefits of multiple calibrators introduced in \citet{2017AJ....153..105R} to the higher frequency ($\ge$5 \,GHz) including weak sources domain, additionally with a user-friendly tool that makes the analysis robust and accessible to a general VLBI user.
The method of sMV is introduced in Sect.~\ref{sec:method}.
In Sect.~\ref{sec:comp}, we compare the results of PR, cMV, and sMV with the same observation data.
In sect.~\ref{sec:discuss}, we discuss the key advantages and observation scheduling of sMV.
Finally, we summarize and look into further application and development in Sect.~\ref{sec:conclusion}.

\section{Method}    \label{sec:method}

The general idea of MultiView is a variant of PR with extra spatial-structure phase correction.
For PR, the phase difference between the target and the only phase calibrator is considered small.
So, the phases estimated through fringe fitting with the calibrator are directly applied to the target.
This works fine if the angular separation between the target and the calibrator is small; however, it is not always possible to find a calibrator close enough to the target.
In the case of a large angular separation, if the residual spatial structure is not corrected, image quality may degrade and there may be systematic errors in position measurement.

Within an area of approximately a few degrees, the residual spatial structure can be approximated with linear gradients, i.e., a phase plane in a 3-dimensional Cartesian space with RA, DEC, and phase as axes.
For MultiView techniques, this plane is estimated through sampling of relative phase differences of several nearby calibrators.
The cMV fits a plane for a group scans on calibrators in each observing cycle, while sMV does not ``fit'' a plane but derives a rotating plane through the whole time series of all scans on calibrators.
This procedure is quite straightforward in the case without phase ambiguity, so we will first introduce the framework in this case, and then discuss how to deal with the tricky ambiguity problem.

\subsection{Case without phase ambiguity}
\label{sec:no_amb}
There are some basic requirements for data to be processed using sMV: more than one phase calibrator near the target source; fast switching between all sources or observing simultaneously (if supported by the VLBI network); at least one (primary) calibrator being strong and compact enough for fringe fitting and having a precise a priori position.
For the calibration of data meeting the above requirements, sMV is a step inserted into the standard astrometric VLBI data calibration flow.
So, the calibration steps in the standard flow should be done before the application of sMV, including amplitude, EOP, ionosphere, and other calibrations.

After the preparations are completed, the sMV procedure begins.
The phase plane is in a 3-dimensional space: RA, DEC, and phase as $X$, $Y$, and $Z$ axes respectively.
The primary calibrator is fixed at the origin (0, 0, 0), and regardless of future iterations, the phase plane will always intersect this point.
Then, fringe fit with the primary calibrator and apply to all other (secondary) calibrators because the phase of the primary calibrator is taken as the zero reference.
On this basis, fringe fitting is done for each secondary calibrator to obtain residual phases of them.
In this subsection, it is assumed that the residual phases never exceed $\pm\pi$ for simplification.
The time series for secondary calibrators can then be expressed as $\{(x, y, \phi, t)_i\}, i=1 \hdots s$, where $(x, y)$ is the position of secondary calibrators relative to the primary calibrator, $\phi$ and $t$ denote residual phase and time respectively, while $s$ is the total number of scans on secondary calibrators.

\begin{figure}[htbp]
  \centering
  \includegraphics[width=0.75\columnwidth]{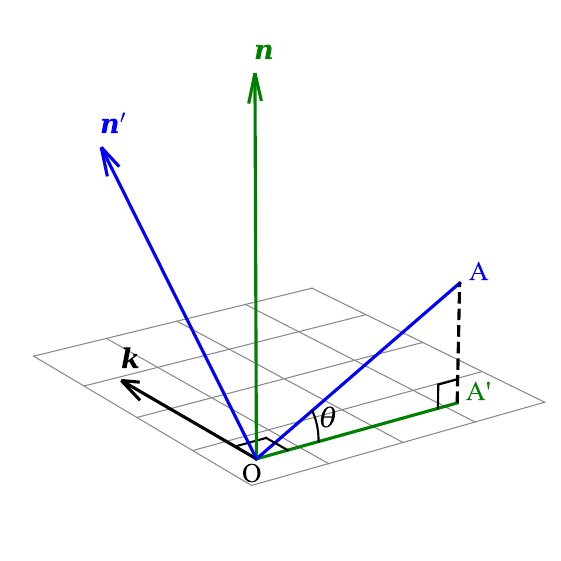}
  \caption{\label{fig:rodrigues}
  Schematic of phase plane rotation.
  O is the original point (0, 0, 0) of the 3-dimensional space, and the gridlines denote the phase plane.
  $\boldsymbol{n}$ is the normal vector of the phase plane to be rotated.
  A is the point that the phase plane is going to pass through.
  A$'$ is the projection of A on the phase plane.
  $\boldsymbol{k}$ is the rotation axis, $\theta$ is the rotation angle, and $\boldsymbol{n}'$ is the normal vector after rotation.
  }
\end{figure}

\begin{figure*}[htbp]
  \centering
  \includegraphics[width=\textwidth]{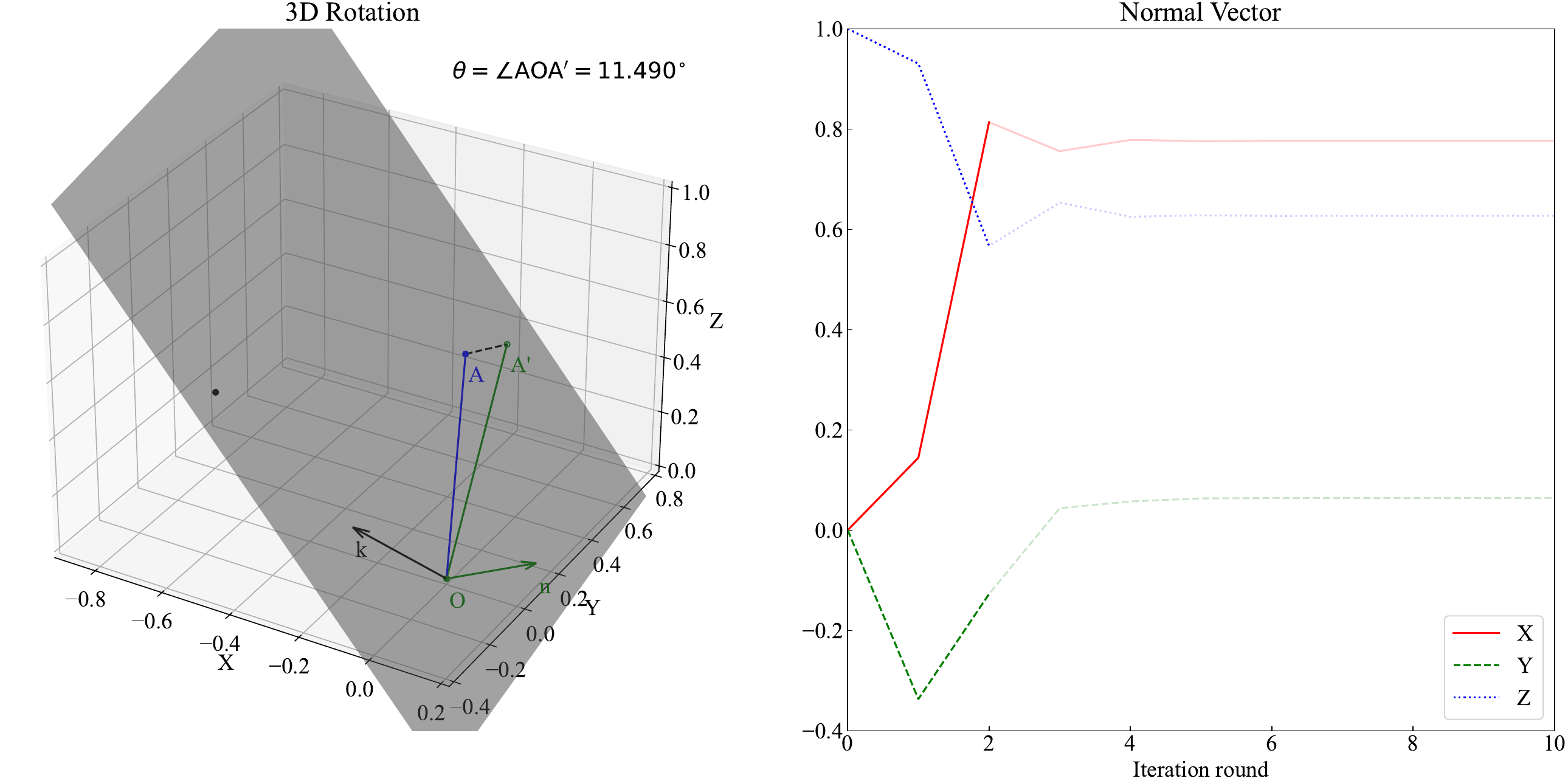}
  \caption{\label{fig:anime}
  Phase plane rotation iteration convergence.
  Notations are the same as those in Fig.~\ref{fig:rodrigues}.
  The changes in the three components of the plane's normal vector are shown on the right.
  An animated version of this figure is available in the online journal.
  The animation steps through the iterations from the initial state and through convergence.
  }
\end{figure*}

Next, an iteration is performed along the time series.
The initial normal vector of the phase plane is set to be (0, 0, 1).
The phase plane is rotated once at each iteration to intersect the new point $(x, y, \phi)_i$.
Since there are already two constraints (the plane must intersect the origin and the new point), only one more constraint is needed to uniquely determine the plane to be rotated to.
The third constraint used here is to minimize the rotation angle, and Fig.~\ref{fig:rodrigues} is a schematic of the rotation.
The cross product of current normal vector and the vector pointing from the origin to the new point ($\boldsymbol{n}\times \boldsymbol{\mathrm{OA}}$) gives the rotation axis $\boldsymbol{k}$, while the rotation angle $\theta$ is the angle between $\boldsymbol{\mathrm{OA}}$ and the phase plane.
The normal vector of the new plane is calculated using the Euler-Rodrigues formula \citep{euler1776nova,JMPA_1840_1_5__380_0,DAI2015144}:
\begin{equation}
    \boldsymbol{n}^{\prime}=\boldsymbol{n}\cos\theta+(\boldsymbol{k}\times\boldsymbol{n})\sin\theta+\boldsymbol{k}(\boldsymbol{k}\cdot\boldsymbol{n})(1-\cos\theta) \ .
\end{equation}
Each rotation reflects the change in phase gradient in the direction of the line connecting the primary calibrator and one secondary calibrator, while iterations in the direction of multiple non-collinear secondary calibrators can converge and reflect the phase gradient on the celestial sphere.
As shown in Fig.~\ref{fig:anime}, over multiple steps during iteration, the phase plane will quickly converge to the vicinity of the ``true'' phase plane.
After converging, only a small rotation angle is needed to track a steadily changing spatial structure.

At the completion of the iteration, we get a phase plane normal vector time series, which is an approximation of the time-varying spatial-structure residual delay.
The next step of sMV is the same as that of cMV: Applying spatial interpolation to obtain the phase correction for the target.
All the sMV steps above should be done for every reference-antenna baseline.
Finally, apply the corrected phases to the target, and then imaging and astrometric fitting can be conducted.

\subsection{Dealing with phase ambiguity}
\label{sec:amb}
Because of the periodicity of the signal, phase ambiguity (signals with phase differences of $2n\pi, n \in \mathbb{Z}$ cannot be distinguished) is a common issue in practice and can introduce large biases, so it is essential to detect and correct it during the sMV iteration.

Ambiguities may occur at any steps during the iteration introduced in Sect.~\ref{sec:no_amb}.
First, we assume no residual phase changes over $\pm2\pi$ between scans, which means that there are three phase wrap options at each step: $+2\pi$, $\pm 0$, or $-2\pi$.
The simplest approach to determine which option is most likely to be ``true'' is to compare the rotation angles needed for the phase plane to rotate from the previous step to them, and the option with the smallest rotation angle should be chosen.
However, this approach is not robust: the measured phases (output of fringe fitting) may include errors caused by low signal-to-noise ratio (SNR), weather condition, or instrument issues, which will result in outliers and lead to misjudgments.
There are many methods for outlier detection in time series processing, e.g., high-pass filtering, interquartile range (IQR), and wavelet transform, but they are not suitable here because it is difficult for them to tell outliers and ``real'' phase wrap apart.

Here we give a robust recursive automatic phase ambiguity detection algorithm with the advantage of being resistant to misjudgments caused by outliers.
At each step in iteration introduced in Sect.~\ref{sec:no_amb}, we need to choose among four options: wrapping by $+2\pi$, no wrap, wrapping by $-2\pi$, and ``this is an outlier''.
If we only compare these four options, no matter what comparison metric we take, it is easy to fall into a local optimum in the case of large noises.
If we can evaluate the impact of the choices made in this step on the subsequent series, then the robustness can be greatly improved:
Outliers usually occur alone or last for a short time, and then the phases will return to the ``normal zone'', leading to a phase plane rotation with a large angular velocity and rapidly changing direction;
while ``real'' phase wraps keep the phase plane rotating in a steady way.
Therefore, if we can find a way to evaluate the subsequent series in (1) stationarity; (2) rotational trend consistency with the preceding series, we can identify outliers and determine the correct phase wrap option.

Let's focus on one step during the iteration.
The preceding normal vector series with length $m$ before this step $\{\boldsymbol{n}_j\}, j=i-m-1 \hdots i-1$ is already determined, so we can easily linearly fit and extrapolate it to get a predicted normal vector $\boldsymbol{n}_{\mathrm{p}}$ at subsequent moments.
However, the case of subsequent series is complex: the phase wrap solutions for subsequent steps remain to be determined, leading to a ternary-tree-shape phase wrap solution set.
That is, each node (includes the root node, corresponding to the $n-1$ step) has three subtrees, therefore there will be $3^n$ leaf nodes for a depth of $n$, and each leaf node represents a unique phase wrap solution for the subsequent series.

The problem now becomes: whether it is possible to find a phase wrap solution with stability and rotational trend consistency that meet the thresholds among all leaf nodes.
If not, the scan at the current step is marked as an outlier;
If possible, among all leaf nodes that meet the thresholds, find the one with the smallest loss and select its corresponding subtree of the root node as the phase wrap solution for this step.
To solve this problem, we need:
\begin{enumerate}
    \item a function for quantitative assessment of stability and rotational trend consistency;
    \item a method for creating and traversing the tree;
    \item a threshold for outlier identification.
\end{enumerate}

The function for stability and rotational trend consistency assessment includes two terms. One is the total rotation angle $\gamma_{\mathrm{tot}}$ accumulated from the root node to the current node, representing the stability;
Another is the angle $\psi_{\mathrm{m}-\mathrm{p}}$ between the rotated normal vector and the predicted normal vector $\boldsymbol{n}_{\mathrm{p}}$, representing the rotational trend consistency.
Combining these two terms with an adjustable weight $w$ will give a reasonable loss function
\begin{equation}
\label{eq:loss}
    f_{\mathrm{loss}} = w\cdot\gamma_{\mathrm{tot}} + (1-w)\cdot\psi_{\mathrm{m}-\mathrm{p}} \ .
\end{equation}

\begin{figure}[htbp]
  \centering
  \includegraphics[width=\columnwidth]{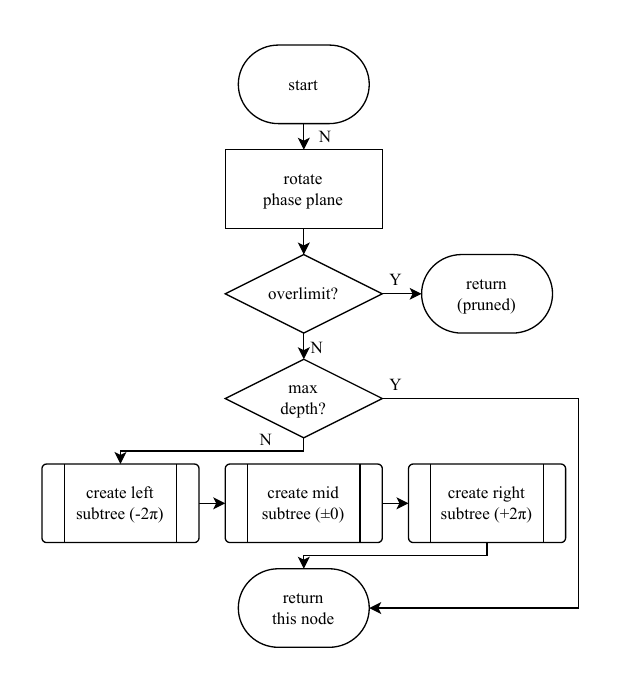}
  \caption{\label{fig:create}
  Flowchart of the recursive function used for creating the ternary tree of phase wrap solutions (pre-order traversal).
  ``Overlimit'' means the pruning threshold is reached, while ``max depth'' means the leaf node has been reached.}
\end{figure}

\begin{figure}[htbp]
  \centering
  \includegraphics[width=\columnwidth]{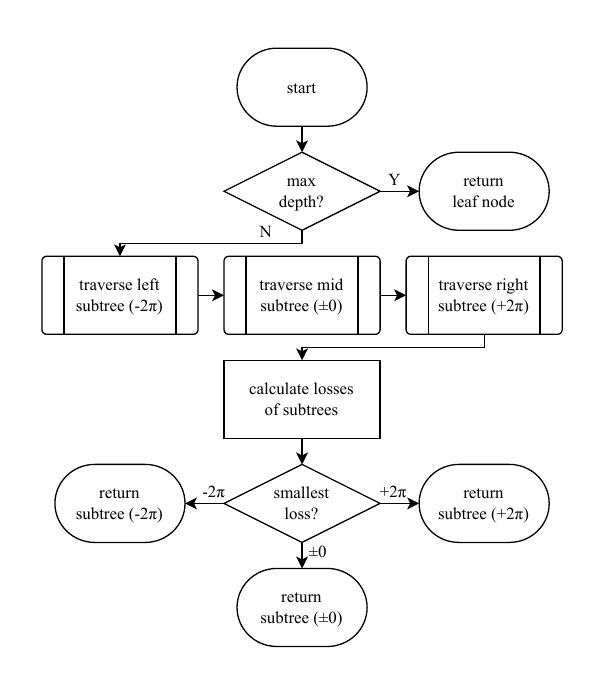}
  \caption{\label{fig:traverse}
  Flowchart of the recursive function used for traversing the ternary tree of phase wrap solutions (post-order traversal).
  ``max depth'' means the leaf node has been reached.
  }
\end{figure}

The most convenient method to create or traverse a tree is recursion.
We need to perform two recursions, the first to create the tree, and the second to find the leaf node with the smallest loss.
Fig.~\ref{fig:create} shows the recursive function we used for creating the tree, and Fig.~\ref{fig:traverse} shows the recursive function we used for traversing the tree.
Note that the time complexity of the recursion is O($3^n$) in the worst case, so the depth should not be too large in practice.

The threshold for outlier identification needs to be carefully selected so that it can distinguish outliers and phase wraps.
There are two thresholds when creating the tree: the maximum allowable plane inclination and rotational angular velocity, which will also help pruning the tree because it is unnecessary to create subtrees for nodes that reach the thresholds.
If none of the $3^n$ leaf nodes can be reached during the tree-creating recursion, the scan at the current step will be marked as an outlier.
There is also a threshold when traversing the tree: the maximum allowable loss value.
Leaf nodes with losses beyond the threshold will be discarded.
If all subtrees of the root node return no leaf nodes, the scan at the current step will be marked as an outlier.

\begin{figure*}[htbp]
  \centering
  \includegraphics[width=\textwidth]{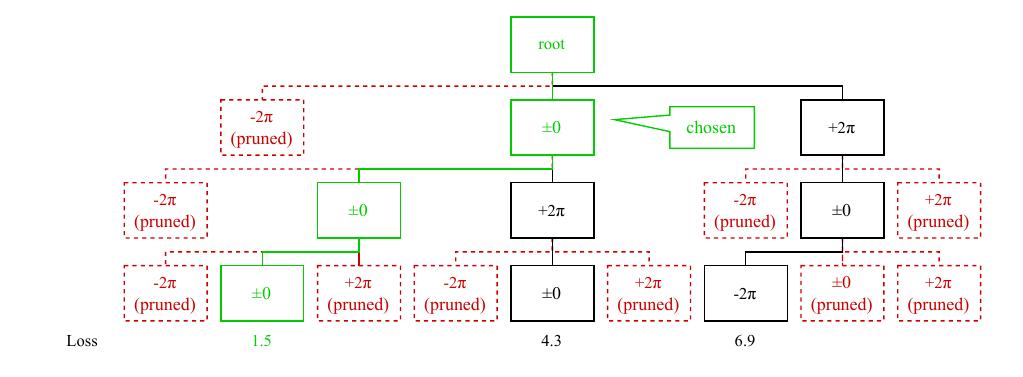}
  \caption{\label{fig:tree}
  Schematic of a ternary tree with a depth $n=3$.
  Red (dashed line) subtrees are pruned.
  Losses of leaf nodes calculated from Eq.~\eqref{eq:loss} are listed below, and the green branch is chosen as the ``best'', so ``$\pm0$'' (at depth $n=1$) is selected as the phase wrap solution for this step.
  }
\end{figure*}

Fig.~\ref{fig:tree} shows an example of how creating, pruning, and traversing a tree works.
In this example, subtrees that reach thresholds are pruned, so that a perfect ternary tree is pruned to have only three leaf nodes, and the subtree at depth $n=1$ with the leaf node that has the smallest loss is selected as the phase wrap solution for this step.
No phase wrap value is accumulated at this step, and the next step will move forward and create and traverse a new tree.

With the above procedure, we can finally determine the phase wrap solution for one step during the iteration, and this procedure is repeated at each step.
Although scans on all secondary calibrators are included in the same iteration process, each secondary calibrator has a separate phase wrap value that accumulates throughout the iteration, and the subsequent scans of each calibrator inherit the phase wrap value accumulated in the preceding iteration.

\subsection{Smoothing}
\label{sec:smo}

Apart from phase ambiguity, there is another potential problem in the iteration: If the secondary calibrators do not agree well with the same phase plane, the normal vector may not converge but oscillate.
This often occurs when the angular separations between calibrators are large or at low elevation, in which cases the residual spatial structure cannot be perfectly approximated as a plane.
Another possible cause is relative position errors between calibrators, which is further discussed in Appendix~\ref{sec:diff}.
A simple way to mitigate the oscillation is to apply smoothing to the normal vector time series.

Here we adopt two filters for smoothing: Kalman filter and low-pass filter.
The Kalman filter is applied during the iteration in real time.
In addition to smoothing the normal vector time series, it also acts as a filter for outliers.
However, it is worth noting that if the smoothing factor is set large, the output of the Kalman filter is significantly delayed compared to the input signal.
So a small smoothing factor is used here, and a low-pass filter is applied after the iteration to improve smoothness.
In practice, this low-pass filter can be implemented as a time-weighted moving average, since the time series are not equally spaced.

\section{Comparison through observation}    \label{sec:comp}

Application to actual observations is the best assessment for sMV.
We have applied sMV to observation data as well as PR and cMV to compare their results.
For example, the Very Long Baseline Array (VLBA) program BZ087.
The program includes observation of 11 radio stars, aiming to measure their parallaxes and proper motions at C band ($\sim$\,4.6\,GHz).
Four calibrators are selected from the Radio Fundamental Catalog \citep[RFC,][]{2025ApJS..276...38P} with VLBI calibrator search engine at Astrogeo\footnote{\url{http://astrogeo.org}} for each target.
The group of sources is arranged in a sequence of ``C1-C2-T-C3-C4'', and the observing cycle is $\sim$\,3 minutes.
An example is given here: radio star V1859 Ori observed on August 19, 2021 (VLBA obs. ID: BZ087B1).
Eight antennas (BR, FD, HN, KP, LA, NL, OV, and PT) were involved in the observation.
Brief information of the target and calibrators is listed in Table~\ref{tab:obs_info}.

\begin{deluxetable*}{lccccc}[htb]
   \tablewidth{\textwidth}
   \setlength{\tabcolsep}{18pt}
   \tablecaption{Brief information of the example observation\label{tab:obs_info}}
   \tablehead{
        Source Name & RA & DEC & $\sigma$ & $S_{\mathrm{unres}}$ & Distance\\
        & (h:m:s) & (d:m:s) & (mas) & (mJy Beam$^{-1}$) & (deg)
   }
   \startdata
        J0517+0648$\star$ & 05:17:51.3442 & +06:48:03.210 & 0.24 & 0.467 (S) & 2.5 \\
        J0519+0848 & 05:19:10.8111 & +08:48:56.734 & 0.21 & 0.253 (S) & 0.93 \\
        J0521+1227 & 05:21:59.7709 & +12:27:05.551 & 0.61 & - & 3.49 \\
        J0532+0732 & 05:32:38.9985 & +07:32:43.345 & 0.22 & 0.222 (C) & 2.8 \\
        \hline
        V1859 Ori & 05:22:54.7927 & +08:58:04.679 & 0.01 & - & - 
   \enddata
   \tablecomments{
        Information of the calibrators is collected from $\texttt{rfc\_2024d}$ (J2000), while that of the target is collected from the SIMBAD database\citep{simbad}, originally from \textit{Gaia} DR3 \citep[][J2016]{2023AA...674A...1G} but propagated to J2000.
        Column $\sigma$ denotes position uncertainty.
        Column $S_{\mathrm{unres}}$ gives the flux density from unresolved components (band in brackets).
        The calibrator with ``$\star$'' is used as the primary calibrator for sMV and calibrator for PR.
        }
\end{deluxetable*}

\begin{figure*}[htbp]
  \centering
  \includegraphics[width=\textwidth]{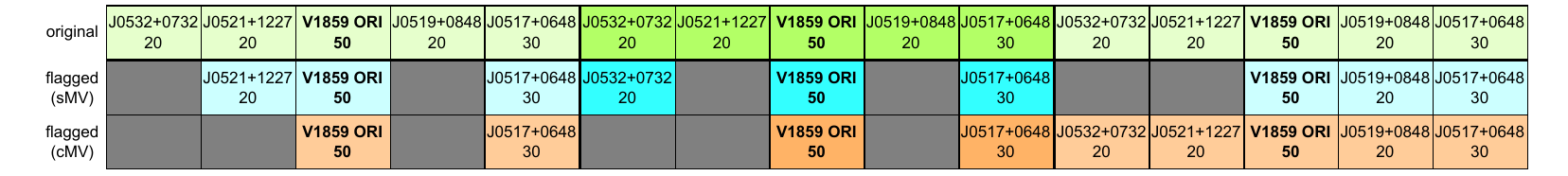}
  \caption{\label{fig:seq}
  Original observing sequence and the flagged sequences.
  Gray cells are flagged to simulate a $1/3$ sampling rate of secondary calibrators.
  Time length in seconds of each scan is shown in each cell.
  }
\end{figure*}

Additionally, we tested a new observing sequence using sMV, as shown in Fig.~\ref{fig:seq}.
Although the actual observing sequence is ``C1-C2-T-C3-C4'', a sequence ``C1-C2-T-C1-T-C3-C1-T-C4'' can be simulated by flagging some scans of calibrators.
This ``flagged'' sequence has 1.4$\times$ on-target time proportion ($\sim$\,50$\%$) than the original one ($\sim$\,36$\%$) if the flagged scans are skipped in observation.
If the time of flagged scans can be added to on-target time, the on-target time proportion will even go up to $\sim$\,64$\%$, 1.8$\times$ that of the original sequence.
To compare sMV with cMV in the case that the calibrator sampling rate is reduced, we also tested the case where only $1/3$ of the cMV phase solutions were used (sequence ``C1-C2-T-C3-C4-C1-T-C1-T''), in which numbers of scans on calibrators are kept the same.

\begin{figure*}[htbp]
    \centering
    \begin{minipage}[b]{0.49\textwidth}
        \centering
        \includegraphics[width=\textwidth]{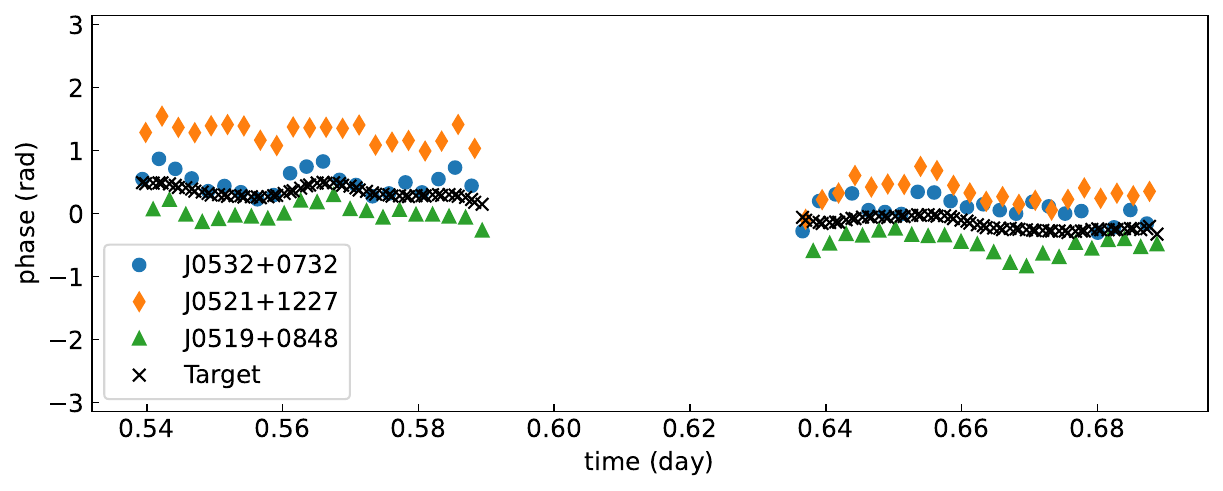}
    \end{minipage}
    \hfill
    \begin{minipage}[b]{0.49\textwidth}
        \centering
        \includegraphics[width=\textwidth]{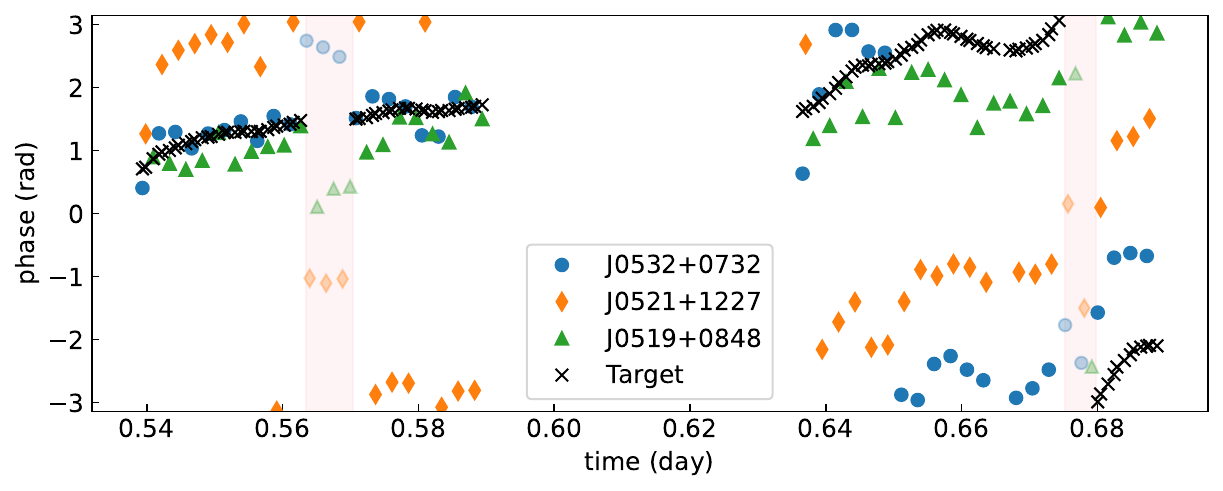}
    \end{minipage}
    \vspace{0.1cm}
    \begin{minipage}[b]{0.49\textwidth}
        \centering
        \includegraphics[width=\textwidth]{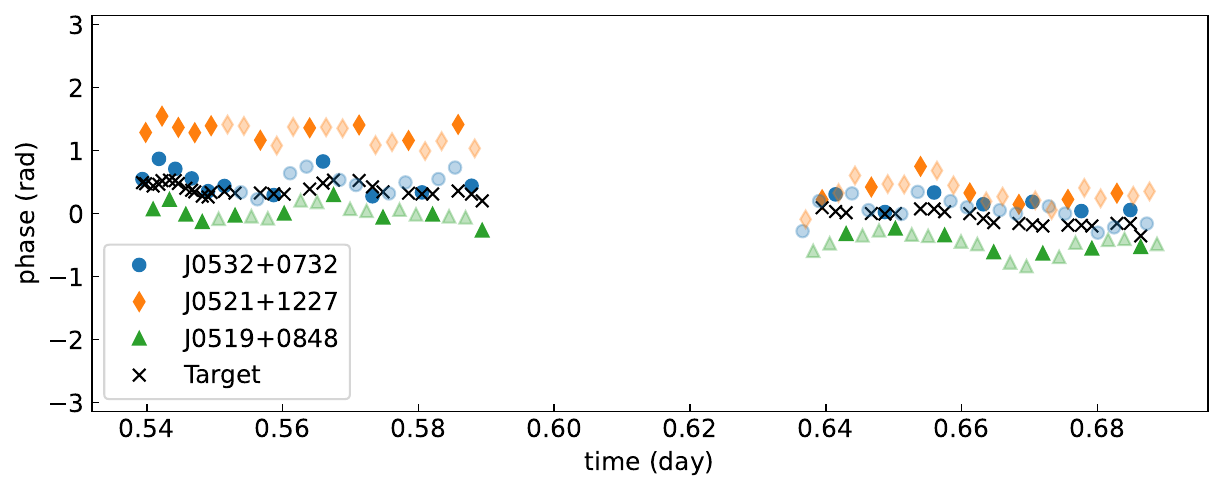}
    \end{minipage}
    \hfill
    \begin{minipage}[b]{0.49\textwidth}
        \centering
        \includegraphics[width=\textwidth]{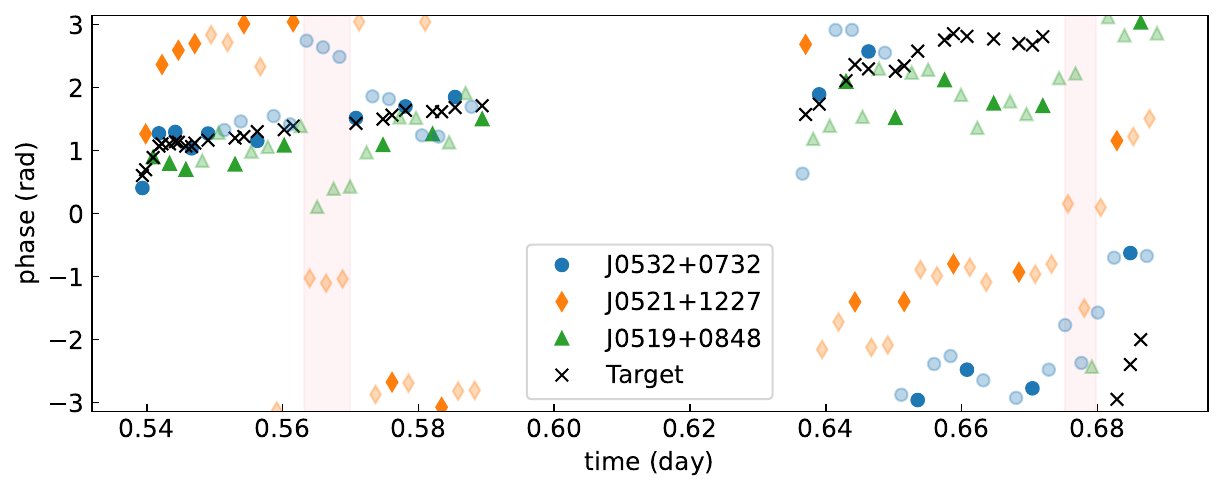}
    \end{minipage}
    \vspace{0.1cm}
    \begin{minipage}[b]{0.49\textwidth}
        \centering
        \includegraphics[width=\textwidth]{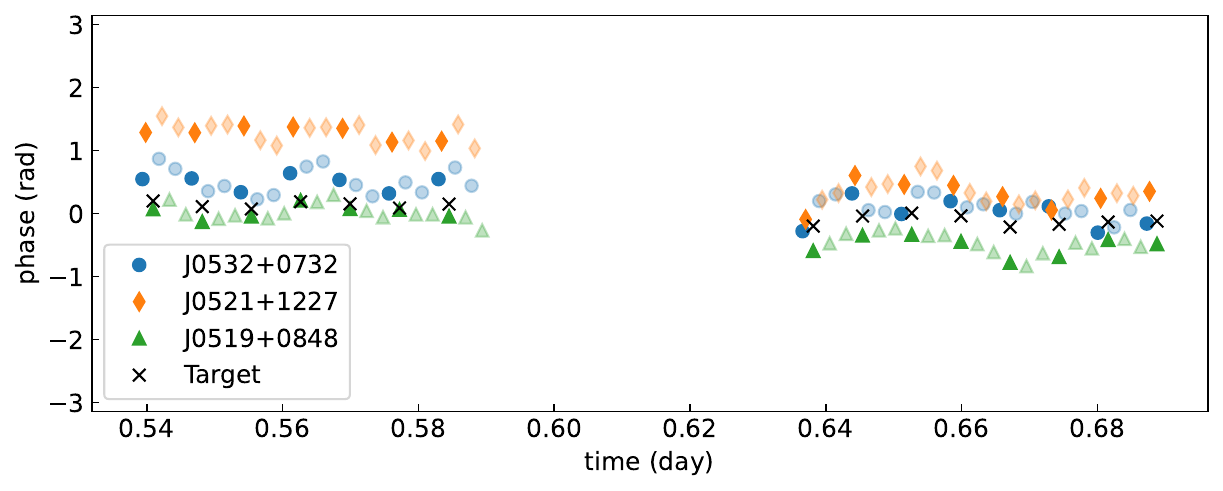}
    \end{minipage}
    \hfill
    \begin{minipage}[b]{0.49\textwidth}
        \centering
        \includegraphics[width=\textwidth]{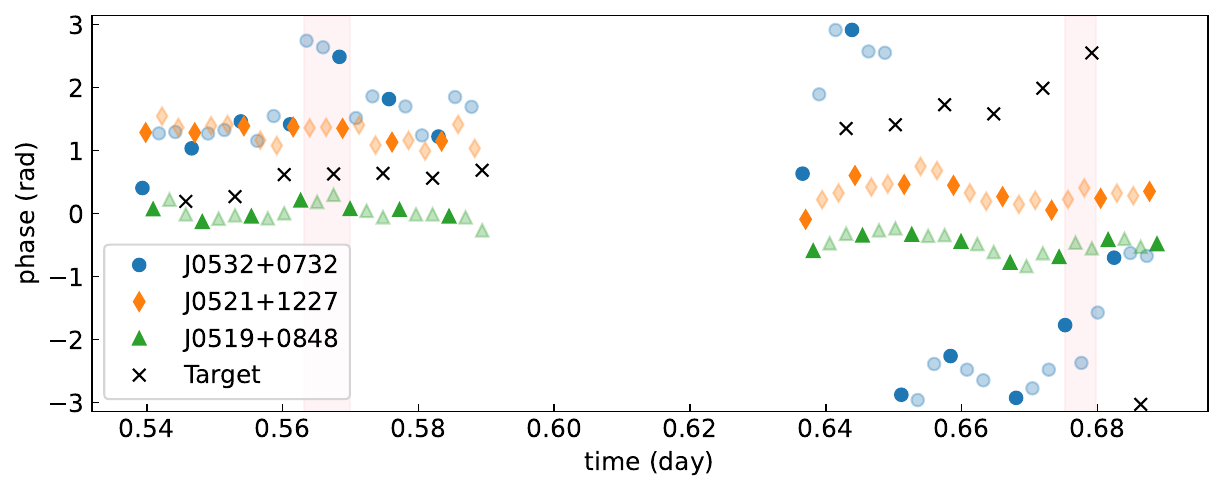}
    \end{minipage}
    \vspace{0.1cm}
    \caption{\label{fig:phase}
    Residual phases in radian of secondary calibrators and estimated phase corrections for the target (V1859 Ori).
    Phases are in radians.
    Left panels: baseline BR-PT; right panels: baseline HN-PT.
    Top panels: original sequences; middle panels: flagged sequences for sMV; bottom panels: flagged sequences for cMV.
    Semitransparent calibrator scans are flagged and not used for sMV or cMV.
    Target scans within time ranges with pink background are not used in imaging.
    }
\end{figure*}

Two examples of secondary calibrator residual phases of are shown in Fig.~\ref{fig:phase}.
When processing the flagged sequence for sMV, a small segment of original sequence is kept at the beginning to help the iteration converge quickly.
In the case of north-south baseline (BR-PT), both antennas had high elevation angles during the whole session, so the residual phase is small, changes slowly and steadily, and there is no phase ambiguity.
The estimated phase corrections for target of two sequences are almost the same.
While in the case of east-west baseline (HN-PT), one antenna (HN) was affected by low elevation angles.
The residual phase changes faster and ambiguity occurs, and the estimated phase correction of both flagged sequences for sMV and cMV slightly loses some of the rapid spatio-temporal jitters in the time series.
This is a foreseeable result, since the sampling rate of secondary calibrators is only $1/3$ of the original sequence.
But the loss is small ($\sim$\,several degrees) and does not have a large impact on interferometric imaging.
It is worth noting that there exist systematic phase biases between sMV and cMV, which are caused by a reference point difference and are discussed in Appendix~\ref{sec:diff}.

\begin{figure*}[htbp]
  \centering
  \includegraphics[width=\textwidth]{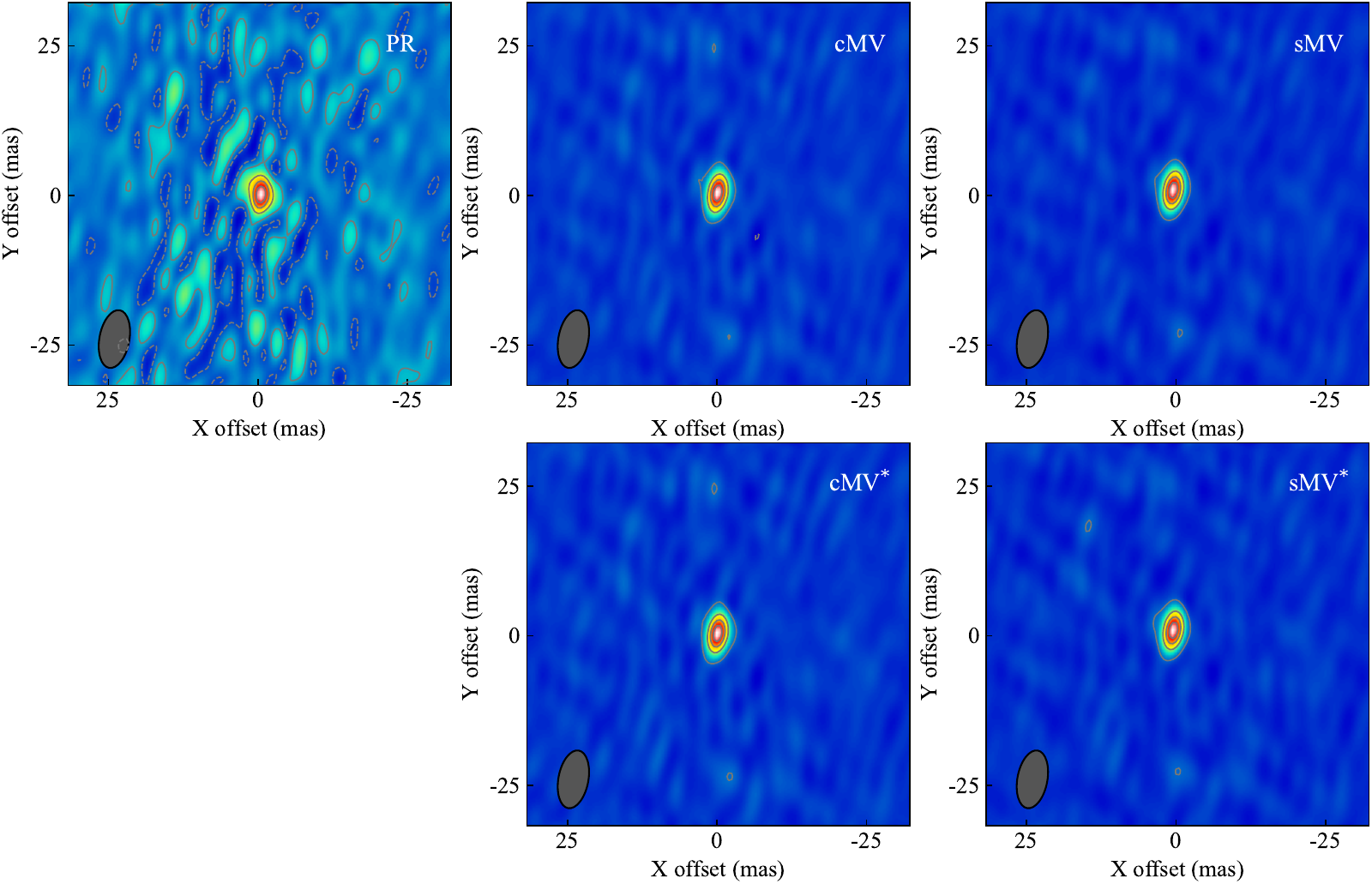}
  \caption{\label{fig:v1859b1}
  Images of V1859 Ori in BZ087B1 calibrated with different techniques.
  cMV$^{\star}$ and sMV$^{\star}$ denote cMV and sMV with flagged sequence respectively.
  Image size is 128$\times$128 with a cellsize of 0.5 mas.
  Contour levels are $[-0.05, 0.05, 0.3, 0.55, 0.8]\,\times$ peak flux density, in which the $-0.05$ level is shown in dashed line.
  The restored beam is shown at the bottom-left corner of each panel.
  }
\end{figure*}

The images given by three calibration techniques (PR, cMV, and sMV with original/flagged sequence) are shown in Fig.~\ref{fig:v1859b1}.
In the PR image, many artifacts indicate that the phase is not calibrated well, reducing the imaging quality; while in MultiView images, these artifacts are mitigated, and the improvement in the reconstruction of point-like source significantly contributes to the fitting of the astrometric center.
This is more pronounced in the east-west direction, as phase ambiguity on east-west baselines is well corrected.
The difference between the original and flagged sequence is hard to tell with the naked eye, proving the feasibility of lowering secondary calibrator sampling rate under good conditions for both techniques.

The astrometric results of different calibration techniques are listed in Table~\ref{tab:obs_res}.
MultiView results are close to each other in flux density and SNR, and are much better than those of PR.
This is as expected, but it's worth noting that the cMV results are achieved after careful manual ambiguity correction, while sMV can do it semi-automatically.
Let's then compare these results with theoretical thermal limit.
The theoretical RMS $\sigma_{\mathrm{thermal}}$ is 42 $\mu$Jy/beam with the eight antennas involved and an on-target time of $\sim$\,36 minutes, calculated using the European VLBI Network (EVN) Observation Planner\footnote{\url{https://planobs.jive.eu}}.
Compared to PR being four times the theoretical value, the MultiView values are quite close to the thermal limit.

\begin{deluxetable*}{lccccccccc}[htb]
   \tablewidth{\textwidth}
   \setlength{\tabcolsep}{5pt}
   \tablecaption{Astrometric results of different calibration techniques of V1859 Ori in BZ087B1\label{tab:obs_res}}
   \tablehead{
        Tech. & $S_{\mathrm{peak}}$ & SNR & RMS & RA & DEC & $\Delta\alpha*$ & $\Delta\delta$ & $\sigma_{\mathrm{RA}}$ & $\sigma_{\mathrm{DEC}}$ \\
         & (mJy Beam$^{-1}$) & & ($\mu$Jy/beam) & (h:m:s) & (d:m:s) & ($\mu$as) & ($\mu$as) & ($\mu$as) & ($\mu$as)
   }
   \startdata
        PR & 3.81\,$\pm$\,0.15 & 25.62 & 165 & 05:22:54.7950173 & $+$08:58:04.479906 & 838 & 789 & 51.7 & 76.5 \\
        cMV & 4.68\,$\pm$\,0.06 & 81.90 & 59 & 05:22:54.7949608 & $+$08:58:04.479117 & - & - & 13.6 & 25.1 \\
        cMV$^{*}$ & 4.64\,$\pm$\,0.06 & 77.85 & 61 & 05:22:54.7949604 & $+$08:58:04.479121 & $-$6 & 4 & 14.5 & 26.4 \\
        sMV & 4.68\,$\pm$\,0.05 & 85.54 & 56 & 05:22:54.7949568 & $+$08:58:04.479273 & $-$59 & 156 & 13.0 & 24.1 \\
        sMV$^{*}$ & 4.62\,$\pm$\,0.06 & 77.61 & 61 & 05:22:54.7949577 & $+$08:58:04.479237 & $-$45 & 120 & 14.6 & 26.5
   \enddata
   \tablecomments{
        $S_{\mathrm{peak}}$ denotes peak flux density.
        $\Delta\alpha*$ and $\Delta\delta$ are position differences relative to cMV (original sequence) results, and $\alpha*$ denotes $\alpha\cos\delta$.
        $\sigma_{\mathrm{RA}}$ and $\sigma_{\mathrm{DEC}}$ are formal uncertainties given by $\mathcal{AIPS}$ \citep{2003ASSL..285..109G} task $\texttt{JMFIT}$.
        }
    \tablenotetext{*}{cMV$^{*}$ and sMV$^{*}$ denote cMV and sMV with flagged sequence respectively.}
\end{deluxetable*}

With the improved SNR, the position formal uncertainties of cMV and sMV are reduced.
Notably, there is an inconsistency in the positions measured by the three calibration techniques.
This could be due to two factors: reference point differences and residual systematic errors.
Reference points of the calibration techniques are different because they make use of the a priori positions of calibrators in different ways, which is discussed in detail in Appendix~\ref{sec:diff}, while in terms of residual systematic errors, consider a case of single baseline: assume wavelength $\lambda=6$\,cm, baseline length $B=4000$\,km, then $\lambda/B\approx 3$\,mas, thus a phase difference of 120$^{\circ}$ results in a position shift of about 1\,mas.
Therefore, to achieve an accuracy of 0.1\,mas level, the residual spatial-structure phase error needs to be corrected to a level below 10$^{\circ}$.

It is not easy to tell reference point differences and residual phase errors apart with a single epoch.
However, they behave differently between epochs: reference point differences should remain constant while other errors change.
Thus, our astrometric verification will come from the comparison of two epochs.
We use another epoch (VLBA obs. ID: BZ087B2) close in time to BZ087B1 as a comparison, which was scheduled the same as B1 and was observed on September 11, 2021.
The data reduction procedure is kept the same between epochs.
If the epochs were very close in time, or if the target was extra-galactic, the repeatability gives a measurement of astrometric errors. 
However, in the case of this paper, the target is a Galactic radio star, and the time between epochs is sufficient for there to have been significant motion.
The solution is to use the fact that cMV is now an established method that is considered to be reliable, thus if cMV and sMV give the same result the second method is also verified.
The images of V1859 Ori in BZ087B2 are shown in Fig.~\ref{fig:v1859b2}, and the astrometric results are shown in Table~\ref{tab:obs_res_b2}.
The position offsets from BZ087B1 to BZ087B2 of each calibration technique include both stellar motion and random errors.
Since cMV is already a mature and reliable technique, we consider its random error to be small, and the position offset of cMV between two epochs is considered to be the ``true'' value of stellar motion.
Therefore, the offsets of PR and sMV subtracted by cMV offsets can be taken as their random errors.
These offsets are listed in Table~\ref{tab:epo_diff}.
PR has random errors of $-$661\,$\mathrm{\mu as}$ in RA direction and $-$278\,$\mathrm{\mu as}$ in DEC direction, indicating large residual phase errors during calibration; while those of sMV are only 9\,$\mathrm{\mu as}$ in RA direction and 25\,$\mathrm{\mu as}$ in DEC direction.
Both results of flagged sequences are also within 1-$\sigma$ range.
This suggests very little residual random errors of sMV and the offset between cMV and sMV is mainly caused by the reference point difference.

\begin{deluxetable*}{lccccccccc}[htb]
   \tablewidth{\textwidth}
   \setlength{\tabcolsep}{5pt}
   \tablecaption{Astrometric results of V1859 Ori in BZ087B2\label{tab:obs_res_b2}}
   \tablehead{
        Tech. & $S_{\mathrm{peak}}$ & SNR & RMS & RA & DEC & $\Delta\alpha*$ & $\Delta\delta$ & $\sigma_{\mathrm{RA}}$ & $\sigma_{\mathrm{DEC}}$ \\
         & (mJy Beam$^{-1}$) & & ($\mu$Jy/beam) & (h:m:s) & (d:m:s) & ($\mu$as) & ($\mu$as) & ($\mu$as) & ($\mu$as)
   }
   \startdata
        PR & 1.26\,$\pm$\,0.03 & 38.25 & 34 & 05:22:54.7950038 & $+$08:58:04.478867 & 177 & 511 & 21.1 & 45.4 \\
        cMV & 1.29\,$\pm$\,0.03 & 42.53 & 30 & 05:22:54.7949919 & $+$08:58:04.478356 & - & - & 15.1 & 36.0 \\
        cMV$^{*}$ & 1.29\,$\pm$\,0.03 & 42.57 & 31 & 05:22:54.7949910 & $+$08:58:04.478331 & $-$13 & $-$25 & 15.0 & 35.6 \\
        sMV & 1.29\,$\pm$\,0.03 & 43.99 & 30 & 05:22:54.7949873 & $+$08:58:04.478487 & $-$68 & 131 & 18.0 & 39.4 \\
        sMV$^{*}$ & 1.28\,$\pm$\,0.03 & 43.34 & 30 & 05:22:54.7949879 & $+$08:58:04.478497 & $-$59 & 141 & 18.2 & 40.1
   \enddata
   \tablecomments{
        All notations are kept the same as Table~\ref{tab:obs_res}.
        }
\end{deluxetable*}

\begin{deluxetable}{lcccccc}[htb]
   \tablewidth{\textwidth}
   \setlength{\tabcolsep}{4pt}
   \tablecaption{Position offsets from BZ087B1 to BZ087B2\label{tab:epo_diff}}
   \tablehead{
        \multirow{2}*{Tech.} & \multicolumn{2}{c}{Offset B2$-$B1} & \multicolumn{2}{c}{Subtract cMV} & \multicolumn{2}{c}{Uncertainty} \\
         & RA & DEC & RA & DEC & RA & DEC \\
         &  ($\mu$as) & ($\mu$as) & ($\mu$as) & ($\mu$as) & ($\mu$as) & ($\mu$as)
   }
   \startdata
        PR & $-$200 & $-$1039 & $-$661 & $-$278 & 72.8 & 121.9 \\
        cMV & 461 & $-$761 & - & - & 28.7 & 61.1 \\
        cMV$^{*}$ & 453 & $-$790 & $-$7 & $-$29 & 29.5 & 62.0 \\
        sMV & 452 & $-$786 & $-$9 & $-$25 & 31.0 & 63.5 \\
        sMV$^{*}$ & 447 & $-$740 & $-$14 & 21 & 32.8 & 66.6 \\
        \textit{Gaia} & 396 & $-$890 & -65 & -129 & 35.0 & 20.1
   \enddata
   \tablecomments{
        Columns 2 and 3 are position offsets from BZ087B1 to BZ087B2, including stellar motion and random errors.
        Columns 4 and 5 are the offsets subtracted by cMV values, which are taken as the ``true'' stellar motion, so that these two columns represent random errors only.
        }
\end{deluxetable}

Additionally, we compared our results to the \textit{Gaia} DR3~\citep{2023AA...674A...1G} prediction.
Although there may be a systematic offset between measured absolute positions of VLBI and \textit{Gaia}, the short-term stellar motion may agree well.
The \textit{Gaia} preliminary measurement predicts stellar motion between August 19 (J2021.63207) and September 11 (J2021.69488), 2021.
Parallax~\citep{2024AA...691A..81D} and proper motion~\citep{2021A&A...649A.124C} corrections for \textit{Gaia} DR3 was applied, and the uncertainty estimation for \textit{Gaia} is based on~\citet[Sect.~5, Eqs. (22)-(24)]{2024MNRAS.529.2062Z}.
We find both MultiView approaches are in much better agreement with the \textit{Gaia} predictions, whereas there is larger discrepancy for PR (i.e. a $\sim$610\,$\mu$as difference for PR and a $\sim$140\,$\mu$as for MultiView).

The differences between the original and flagged sequences are very small: measured flux densities and SNR of both sequences are almost the same, and the position difference is also small, for both sMV and cMV and in both B1 and B2.
It is quite satisfying to achieve comparable results with only $\sim$\,1/3 of sampling rate of secondary calibrators.
This proves the potential of MultiView techniques in improving on-target time proportion.
As a rough estimate, with the time of flagged scans added to on-target time (1.8$\times$), the SNR would go up to $\sim$\,103 for B1 and $\sim$\,58 for B2.

\section{Discussion} \label{sec:discuss}

\subsection{Advantages of sMV for mid and high frequencies}
Although in principle, in most cases cMV and sMV should provide similar calibration quality for the same data, there are several key advantages of sMV in comparison with the conventional approach, from scheduling to calibration:

\subsubsection*{Automaticity}
sMV comprises of a robust recursive automatic phase ambiguity detection algorithm, which has the advantage of being resistant to misjudgments caused by outliers.
The iteration along the time axis enables sMV to identify outliers and phase wraps in terms of the continuity and consistency of the time series.
The iteration procedure usually does not need any manual intervention, reducing human workload and the potential errors due to subjective judgment.
Since the phase plane of each cMV cycle is fitted independently, fully automatic and accurate identification of outliers and phase wraps currently requires user expertise, and there is no publicly released tool.
The phase ambiguity correction is a fundamental requirement for the astrometric verification presented in this paper.

\subsubsection*{Coherence time limit}
The interferometer coherence time is the key limitation to the application of MultiView at higher frequencies ($\ge$\,8 GHz).
No matter how the calibrator sampling rate is reduced, in order to ensure the quasi-simultaneity of calibrator scans, cMV still requires observing all calibrators in the same cycle.
In the case of Sect.~\ref{sec:comp}, the total time of a cMV cycle is 140 seconds plus slewing time of about 70 seconds, while the time interval between two primary calibrator scans of sMV is 100 seconds plus slewing time of about 40 seconds if the observing cycle is shortened.
This makes it possible to apply MultiView at higher frequencies, which is future work..

\subsubsection*{Flexibility}
sMV provides flexibility in observing sequence and calibrator number.
In principle, any number of calibrators ($\ge$\,2) and targets can be combined in any order to meet different needs.
It is also possible to use different subsequences in one session, for example, use a lower secondary calibrator sampling rate at high elevation and a higher one at low elevation to maximize efficiency.

\subsubsection*{Less slew time}
It is unnecessary to shorten cycle lengths under good conditions (e.g., at mid frequencies).
In this case, the length of the scans on target can be extended so that the antennas will switch between sources less frequently.
This is important not only for time-saving -- for large antennas, frequent slewing is potentially damaging to their rotating mechanism.

\subsection{Observation scheduling}
An important prerequisite for good sMV phase calibration is a well-scheduled observation.
Scheduling for different targets should be carefully designed case by case.
Factors such as the observing frequency, the VLBI network used, the target source, calibrators nearby, etc. all have a great influence on scheduling, but there are still some general laws.

The first question to answer is how many calibrators are required.
If you can find two nearby calibrators approximately in a line with the target, 1-dimensional phase interpolation can be applied;
If three calibrators around the target within $\sim 2^{\circ}$ can be found, 2-dimensional MultiView is a good choice;
If the calibrators are far from the target, it is more robust to add another calibrator for distinguishing phase ambiguity.

The next question is about the sequence.
A ``full'' sampling rate of secondary calibrators (e.g., ``C1-C2-T-C3-C4-\dots'' as cMV) provides most details of rapid spatio-temporal jitters in the residual phase time series, but sacrifices much in terms of on-target time.
If the frequencies and/or the elevation are not low, or if the session is short, or if the baselines are not particularly long, the atmospheric spatial structure may not change fast.
In this case, a reduced sampling rate can be used.
As discussed above, it is also possible to adopt a hybrid sequence: a higher sampling rate at the beginning and end of the session and a lower sampling rate in the middle.

The above discussion is for traditional VLBI facilities with a single-beam antenna at each site.
With advanced or next-generation facilities, there are more scheduling possibilities.
For example, for dual-beam systems, duty can be divided between beams.
It is theoretically possible to track the target with one beam, and switch between calibrators with another beam.
It is even possible to track all sources simultaneously with multi-beam systems, which avoids interpolation in the time domain.

For co-located antennas, i.e., ``cluster-cluster'' mode \citep{1997VA.....41..213R}, the case is similar to multi-beam systems but offers more flexibilities: sub-array configuration at each site, cooperation between large-but-slow and small-but-fast antennas, and so on.
This provides the opportunity to exploit the astrometric performance potential of the VLBI facilities and broaden the applicable frequency range.

\subsection{In conjunction with geodetic blocks} \label{sec:geo}

Inserting geodetic blocks into regular PR sessions is also an effective method for atmospheric correction~\citep{2009ApJ...693..397R,2022PASP..134l3001R}.
Geodetic blocks consist of observations of a large number of calibrators across the sky,
using which clock and residual atmospheric path delays can be fitted from the group delays.
One block typically lasts about half an hour and can be interleaved in PR sessions.
Compared with MultiView techniques, geodetic blocks require less observing time and are not bothered by the phase ambiguity problem.
Non-dispersive tropospheric delays versus elevation are fitted to a mapping function, and the typical tropospheric errors are reduced from $\sim$3\,cm to $\sim$1\,cm, reducing the phase residuals, which can allow for astrometric precisions as high as 20\,$\mu$as at $\sim$22\,GHz.

MultiView is expected to provide much better precision and at a greater range of frequencies.
Geoblocks are only really applicable at a higher frequency (e.g., 24\,GHz, \citealt{2022PASP..134l3001R}), but may be important in the extension of MultiView techniques to these frequencies, where phase ambiguities will become even more challenging.

\section{Summary and future outlook} \label{sec:conclusion}
In this paper, we present a new approach of MultiView, serial MultiView, targeted at overcoming several shortcomings of the conventional MultiView approach and provide a user-friendly tool for it.
This new approach yields improvements in observing cycle length, and provides highly automated phase ambiguity correction.

Comparison through calibrating the same data with PR, cMV, and sMV proves that sMV can achieve significantly improved image quality and astrometric parameters than that of PR, and handles ambiguity problem more automatically than cMV.
By comparing the results of two epochs, we can deduce that the difference between sMV and cMV in the individual epochs mainly comes from the different virtual reference points, whilst the astrometry they provide agree very well with each other.

A test for a lower secondary calibrator sampling rate yields a good result, proving that MultiView techniques have great potential to increase on-target time proportion without reducing calibration quality significantly.
Another potential of sMV is in shortening the observing cycle, which makes it possible for MultiView to be applied at a higher frequency.
We are going to try different observing sequences, e.g., ``C1-T-C2-C1-T-C3-C1-T-C4'', through test observations in the near future.
We also plan to test sMV at a higher frequency, e.g., X or K band.

We welcome all VLBI observers to try sMV.
The tool we provide makes it very simple to insert sMV into your data reduction flow, and it can significantly improve your imaging quality and astrometric results.
It is also valuable for astrophysical research: artifacts in images decrease with well-calibrated phases.
We particularly look forward to a wide application of sMV with next-generation facilities, e.g., the Square Kilometre Array (SKA, \citet{2009IEEEP..97.1482D}) and the next-generation Very Large Array (ngVLA, \citet{2015arXiv151006438C}).


\section*{Acknowledgments}

We want to thank the referee for the comments and suggestions which were helpful in improving this paper.
J.Z. would also like to thank Dr. Leonid Petrov and Dr. Hao Ding for their valuable suggestions.
S.X. thanks Dr. Mark Reid and Dr. Lucas Hyland for sharing the technical details of inverse MultiView.

This work is supported by the National Natural Science Foundation of China (NSFC) under grant Ns. U2031212, the National Key R\&D Program of China (No. 2024YFA1611500), and the Strategic Priority Research Program of the Chinese Academy of Sciences, Grant No. XDA0350205.

The Python code of the sMV procedure can be found at \url{https://github.com/FrdCHK/serial-MultiView}, and a ``frozen'' version used in this paper is available at Zenodo \citep{jingdong_zhang_2025_15030432}: \url{https://doi.org/10.5281/zenodo.15030432}.
We also provide a pipeline for calibration, imaging, and astrometry, which can be found at the same URL.
The code and examples can also be obtained by contacting the authors upon request.

VLBA data used in this paper can be downloaded from \url{https://data.nrao.edu}.
This work has made use of data from the European Space Agency (ESA) mission
{\it Gaia} (\url{https://www.cosmos.esa.int/gaia}), processed by the {\it Gaia}
Data Processing and Analysis Consortium (DPAC,
\url{https://www.cosmos.esa.int/web/gaia/dpac/consortium}). Funding for the DPAC
has been provided by national institutions, in particular the institutions
participating in the {\it Gaia} Multilateral Agreement.
Python code for \textit{Gaia} DR3 parallax zero-point correction used in this paper can be found at \url{https://github.com/yedings/Parallax-bias-correction-in-the-Galactic-plane}.

The RFC can be accessed through \url{https://doi.org/10.25966/dhrk-zh08}.
This work has also made use of the SIMBAD database, operated at CDS, Strasbourg, France.


%

\vspace{5mm}
\facilities{VLBA, \textit{Gaia}}


\software{Numpy \citep{harris2020array}, Pandas \citep{2022zndo...3509134T}, Scipy \citep{2020NatMe..17..261V}, Matplotlib \citep{2007CSE.....9...90H}, Astropy \citep{astropy2013,astropy2018,astropy2022}, $\mathcal{AIPS}$ \citep{2003ASSL..285..109G}, ParselTongue \citep{2006ASPC..351..497K}.
          }


\appendix

\section{BZ087B2 images}
\label{sec:b2}

\begin{figure*}[htbp]
  \centering
  \includegraphics[width=\textwidth]{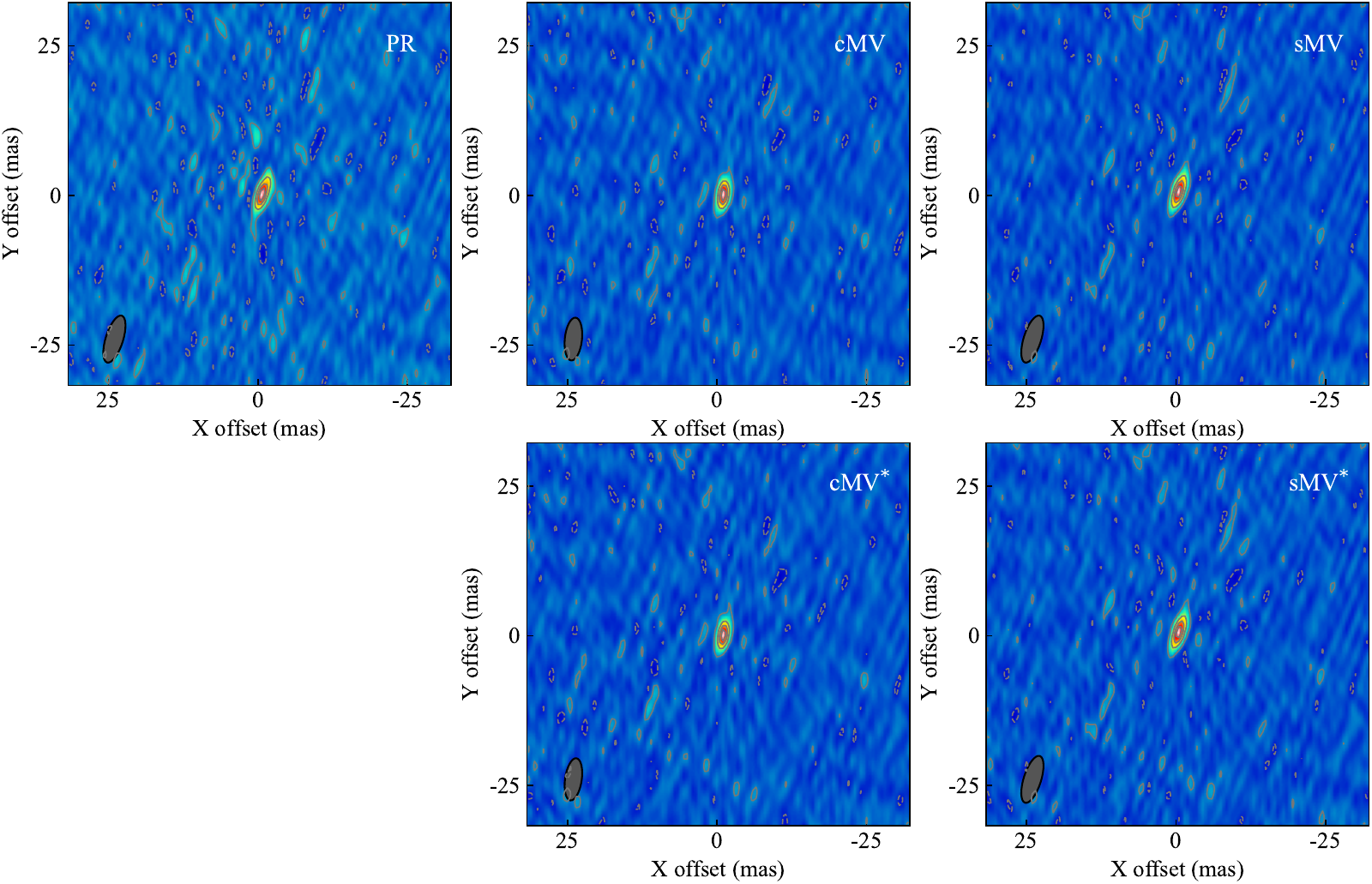}
  \caption{\label{fig:v1859b2}
  Images of V1859 Ori in BZ087B2 calibrated with different techniques.
  All notations are kept the same as Fig.~\ref{fig:v1859b1}.
  }
\end{figure*}

The images of V1859 Ori in BZ087B2 are shown in Fig.~\ref{fig:v1859b2}.

\section{Reference point differences between cMV and sMV} \label{sec:diff}
Although the two approaches are very similar in principle, the reference point difference between them is worth discussing.
Suppose the input data are the same: one calibrator is fixed at the origin in the 3D space, and points in the space represent residual phases of the scans on the other calibrators.
The cMV fits a free phase plane with no additional constraints (in a form of $z=ax+by+c$), while sMV is equivalent to fitting a phase plane with a constraint that it must intersect the origin ($z=ax+by$).
Considering the case where there are no additional error terms and the SNR of fringe fitting is high, sMV is basically the same as cMV.
However, if there are non-negligible error terms in fringe fitting, things will be different.
Random errors can be mitigated through smoothing, but systematic errors, for example, relative position errors, will cause systematic phase biases in fringe fitting.
Source structure and core shift may also cause similar systematic errors.
The impact of these systematic errors on the measured target position is analyzed below.

To simplify the mathematical form for easier discussion, we assume all calibrators are arranged in a straight line, and then the problem degenerates into fitting a straight line in a 2D plane ($y=ax+b$ for cMV and $y=ax$ for sMV).
The input data is a set of points $\{(x_i, y_i)\}, i=1\hdots n$ that are independent and have the same uncertainty, and the mean value of $\{x_i\}$ and $\{y_i\}$ are $\mu_x=\frac{1}{n}\sum_{i=1}^{n}x_i$ and $\mu_y=\frac{1}{n}\sum_{i=1}^{n}y_i$ respectively.

For cMV, the slope $a$ and the y-intercept $b$ can be estimated through
\begin{equation}
\label{eq:cmv_lsq}
\begin{aligned}
    a&=\frac{\sum_{i=1}^{n}(x_i-\mu_x)y_i}{\sum_{i=1}^{n}(x_i-\mu_x)^2} \ ,\\
    b&=\mu_y-a\mu_x \ .
\end{aligned}
\end{equation}
The impact of a deviation of one data point $(x_j, y_j), j=1\hdots n$ on $a$ and $b$ can be estimated through differentiating $a$ and $b$ with respect to $y_j$:
\begin{equation}
\label{eq:cmv_partial}
\begin{aligned}
    \frac{\partial a}{\partial y_j}&=\frac{x_j-\mu_x}{\sum_{i=1}^{n}(x_i-\mu_x)^2} \ ,\\
    \frac{\partial b}{\partial y_j}&=\frac{1}{n}-\frac{\mu_x(x_j-\mu_x)}{\sum_{i=1}^{n}(x_i-\mu_x)^2} \ .
\end{aligned}
\end{equation}
If we calculate the function value $y_{\mathrm{T}}$ at $x=x_{\mathrm{T}}, x_{\mathrm{T}}\in \mathbb{R}$ using function $y=ax+b$, the impact of the deviation of $(x_j, y_j)$ on $y_{\mathrm{T}}$ will be
\begin{equation}
\label{eq:cmv_yc}
    \frac{\partial y_{\mathrm{T}}}{\partial y_j}=\frac{1}{n}+\frac{(x_{\mathrm{T}}-\mu_x)(x_j-\mu_x)}{\sum_{i=1}^{n}(x_i-\mu_x)^2} \ .
\end{equation}
If $x_{\mathrm{T}}=\mu_x$, the second term of Eq.~\eqref{eq:cmv_yc} being always 0 means that at the deviations of all input data points contribute equally to the deviation of the function value at $x=\mu_x$.
In other words, for cMV, at the mean coordinates of all calibrators, the systematic error of the function value comes from the arithmetic mean of the systematic errors (such as relative position errors) of each calibrator, which makes intuitive sense.

For sMV, the slope $a$ can be estimated through
\begin{equation}
\label{eq:smv_lsq}
    a=\frac{\sum_{i=1}^{n}x_iy_i}{\sum_{i=1}^{n}x_i^2} \ .
\end{equation}
The impact of a deviation of one data point $(x_j, y_j)$ on $a$ can be estimated through differentiating $a$ with respect to $y_j$:
\begin{equation}
\label{eq:smv_partial}
    \frac{\partial a}{\partial y_j}=\frac{x_j}{\sum_{i=1}^{n}x_i^2} \ .
\end{equation}
Similarly, calculate the function value $y_{\mathrm{T}}$ at $x=x_{\mathrm{T}}$ using function $y=ax$, the impact of the deviation of $(x_j, y_j)$ on $y_{\mathrm{T}}$ will be
\begin{equation}
\label{eq:smv_yc}
    \frac{\partial y_{\mathrm{T}}}{\partial y_j}=\frac{x_{\mathrm{T}}x_j}{\sum_{i=1}^{n}x_i^2} \ ,
\end{equation}
which is different from Eq.~\eqref{eq:cmv_yc}, and this is why there is a reference point difference between cMV and sMV.
Meanwhile, the reference point of single-calibrator PR is the only calibrator itself.
This reference point difference usually stays the same between sessions, so it will not affect parallax or proper motion measurement, unless the source structures of the calibrators change very quickly.


\bibliography{main}{}

\begin{thebibliography}{}
\expandafter\ifx\csname natexlab\endcsname\relax\def\natexlab#1{#1}\fi
\providecommand{\url}[1]{\href{#1}{#1}}
\providecommand{\dodoi}[1]{doi:~\href{http://doi.org/#1}{\nolinkurl{#1}}}
\providecommand{\doeprint}[1]{\href{http://ascl.net/#1}{\nolinkurl{http://ascl.net/#1}}}
\providecommand{\doarXiv}[1]{\href{https://arxiv.org/abs/#1}{\nolinkurl{https://arxiv.org/abs/#1}}}

\bibitem[{{Astropy Collaboration} {et~al.}(2013){Astropy Collaboration},
  {Robitaille}, {Tollerud}, {Greenfield}, {Droettboom}, {Bray}, {Aldcroft},
  {Davis}, {Ginsburg}, {Price-Whelan}, {Kerzendorf}, {Conley}, {Crighton},
  {Barbary}, {Muna}, {Ferguson}, {Grollier}, {Parikh}, {Nair}, {Unther},
  {Deil}, {Woillez}, {Conseil}, {Kramer}, {Turner}, {Singer}, {Fox}, {Weaver},
  {Zabalza}, {Edwards}, {Azalee Bostroem}, {Burke}, {Casey}, {Crawford},
  {Dencheva}, {Ely}, {Jenness}, {Labrie}, {Lim}, {Pierfederici}, {Pontzen},
  {Ptak}, {Refsdal}, {Servillat}, \& {Streicher}}]{astropy2013}
{Astropy Collaboration}, {Robitaille}, T.~P., {Tollerud}, E.~J., {et~al.} 2013,
  \aap, 558, A33, \dodoi{10.1051/0004-6361/201322068}

\bibitem[{{Astropy Collaboration} {et~al.}(2018){Astropy Collaboration},
  {Price-Whelan}, {Sip{\H{o}}cz}, {G{\"u}nther}, {Lim}, {Crawford}, {Conseil},
  {Shupe}, {Craig}, {Dencheva}, {Ginsburg}, {VanderPlas}, {Bradley},
  {P{\'e}rez-Su{\'a}rez}, {de Val-Borro}, {Aldcroft}, {Cruz}, {Robitaille},
  {Tollerud}, {Ardelean}, {Babej}, {Bach}, {Bachetti}, {Bakanov}, {Bamford},
  {Barentsen}, {Barmby}, {Baumbach}, {Berry}, {Biscani}, {Boquien}, {Bostroem},
  {Bouma}, {Brammer}, {Bray}, {Breytenbach}, {Buddelmeijer}, {Burke},
  {Calderone}, {Cano Rodr{\'\i}guez}, {Cara}, {Cardoso}, {Cheedella}, {Copin},
  {Corrales}, {Crichton}, {D'Avella}, {Deil}, {Depagne}, {Dietrich}, {Donath},
  {Droettboom}, {Earl}, {Erben}, {Fabbro}, {Ferreira}, {Finethy}, {Fox},
  {Garrison}, {Gibbons}, {Goldstein}, {Gommers}, {Greco}, {Greenfield},
  {Groener}, {Grollier}, {Hagen}, {Hirst}, {Homeier}, {Horton}, {Hosseinzadeh},
  {Hu}, {Hunkeler}, {Ivezi{\'c}}, {Jain}, {Jenness}, {Kanarek}, {Kendrew},
  {Kern}, {Kerzendorf}, {Khvalko}, {King}, {Kirkby}, {Kulkarni}, {Kumar},
  {Lee}, {Lenz}, {Littlefair}, {Ma}, {Macleod}, {Mastropietro}, {McCully},
  {Montagnac}, {Morris}, {Mueller}, {Mumford}, {Muna}, {Murphy}, {Nelson},
  {Nguyen}, {Ninan}, {N{\"o}the}, {Ogaz}, {Oh}, {Parejko}, {Parley}, {Pascual},
  {Patil}, {Patil}, {Plunkett}, {Prochaska}, {Rastogi}, {Reddy Janga},
  {Sabater}, {Sakurikar}, {Seifert}, {Sherbert}, {Sherwood-Taylor}, {Shih},
  {Sick}, {Silbiger}, {Singanamalla}, {Singer}, {Sladen}, {Sooley},
  {Sornarajah}, {Streicher}, {Teuben}, {Thomas}, {Tremblay}, {Turner},
  {Terr{\'o}n}, {van Kerkwijk}, {de la Vega}, {Watkins}, {Weaver}, {Whitmore},
  {Woillez}, {Zabalza}, \& {Astropy Contributors}}]{astropy2018}
{Astropy Collaboration}, {Price-Whelan}, A.~M., {Sip{\H{o}}cz}, B.~M., {et~al.}
  2018, \aj, 156, 123, \dodoi{10.3847/1538-3881/aabc4f}

\bibitem[{{Astropy Collaboration} {et~al.}(2022){Astropy Collaboration},
  {Price-Whelan}, {Lim}, {Earl}, {Starkman}, {Bradley}, {Shupe}, {Patil},
  {Corrales}, {Brasseur}, {N{\"o}the}, {Donath}, {Tollerud}, {Morris},
  {Ginsburg}, {Vaher}, {Weaver}, {Tocknell}, {Jamieson}, {van Kerkwijk},
  {Robitaille}, {Merry}, {Bachetti}, {G{\"u}nther}, {Aldcroft},
  {Alvarado-Montes}, {Archibald}, {B{\'o}di}, {Bapat}, {Barentsen},
  {Baz{\'a}n}, {Biswas}, {Boquien}, {Burke}, {Cara}, {Cara}, {Conroy},
  {Conseil}, {Craig}, {Cross}, {Cruz}, {D'Eugenio}, {Dencheva}, {Devillepoix},
  {Dietrich}, {Eigenbrot}, {Erben}, {Ferreira}, {Foreman-Mackey}, {Fox},
  {Freij}, {Garg}, {Geda}, {Glattly}, {Gondhalekar}, {Gordon}, {Grant},
  {Greenfield}, {Groener}, {Guest}, {Gurovich}, {Handberg}, {Hart},
  {Hatfield-Dodds}, {Homeier}, {Hosseinzadeh}, {Jenness}, {Jones}, {Joseph},
  {Kalmbach}, {Karamehmetoglu}, {Ka{\l}uszy{\'n}ski}, {Kelley}, {Kern},
  {Kerzendorf}, {Koch}, {Kulumani}, {Lee}, {Ly}, {Ma}, {MacBride}, {Maljaars},
  {Muna}, {Murphy}, {Norman}, {O'Steen}, {Oman}, {Pacifici}, {Pascual},
  {Pascual-Granado}, {Patil}, {Perren}, {Pickering}, {Rastogi}, {Roulston},
  {Ryan}, {Rykoff}, {Sabater}, {Sakurikar}, {Salgado}, {Sanghi}, {Saunders},
  {Savchenko}, {Schwardt}, {Seifert-Eckert}, {Shih}, {Jain}, {Shukla}, {Sick},
  {Simpson}, {Singanamalla}, {Singer}, {Singhal}, {Sinha}, {Sip{\H{o}}cz},
  {Spitler}, {Stansby}, {Streicher}, {{\v{S}}umak}, {Swinbank}, {Taranu},
  {Tewary}, {Tremblay}, {Val-Borro}, {Van Kooten}, {Vasovi{\'c}}, {Verma}, {de
  Miranda Cardoso}, {Williams}, {Wilson}, {Winkel}, {Wood-Vasey}, {Xue},
  {Yoachim}, {Zhang}, {Zonca}, \& {Astropy Project Contributors}}]{astropy2022}
{Astropy Collaboration}, {Price-Whelan}, A.~M., {Lim}, P.~L., {et~al.} 2022,
  \apj, 935, 167, \dodoi{10.3847/1538-4357/ac7c74}

\bibitem[{{Beasley} \& {Conway}(1995)}]{1995ASPC...82..327B}
{Beasley}, A.~J., \& {Conway}, J.~E. 1995, in Astronomical Society of the
  Pacific Conference Series, Vol.~82, Very Long Baseline Interferometry and the
  VLBA, ed. J.~A. {Zensus}, P.~J. {Diamond}, \& P.~J. {Napier}, 327

\bibitem[{{Cantat-Gaudin} \& {Brandt}(2021)}]{2021A&A...649A.124C}
{Cantat-Gaudin}, T., \& {Brandt}, T.~D. 2021, \aap, 649, A124,
  \dodoi{10.1051/0004-6361/202140807}

\bibitem[{{Carilli} {et~al.}(2015){Carilli}, {McKinnon}, {Ott}, {Beasley},
  {Isella}, {Murphy}, {Leroy}, {Casey}, {Moullet}, {Lacy}, {Hodge}, {Bower},
  {Demorest}, {Hull}, {Hughes}, {di Francesco}, {Narayanan}, {Kent}, {Clark},
  \& {Butler}}]{2015arXiv151006438C}
{Carilli}, C.~L., {McKinnon}, M., {Ott}, J., {et~al.} 2015, arXiv e-prints,
  arXiv:1510.06438, \dodoi{10.48550/arXiv.1510.06438}

\bibitem[{Dai(2015)}]{DAI2015144}
Dai, J.~S. 2015, Mechanism and Machine Theory, 92, 144,
  \dodoi{https://doi.org/10.1016/j.mechmachtheory.2015.03.004}

\bibitem[{{Dewdney} {et~al.}(2009){Dewdney}, {Hall}, {Schilizzi}, \&
  {Lazio}}]{2009IEEEP..97.1482D}
{Dewdney}, P.~E., {Hall}, P.~J., {Schilizzi}, R.~T., \& {Lazio}, T.~J.~L.~W.
  2009, IEEE Proceedings, 97, 1482, \dodoi{10.1109/JPROC.2009.2021005}

\bibitem[{{Ding} {et~al.}(2023){Ding}, {Deller}, {Stappers}, {Lazio}, {Kaplan},
  {Chatterjee}, {Brisken}, {Cordes}, {Freire}, {Fonseca}, {Stairs},
  {Guillemot}, {Lyne}, {Cognard}, {Reardon}, \&
  {Theureau}}]{2023MNRAS.519.4982D}
{Ding}, H., {Deller}, A.~T., {Stappers}, B.~W., {et~al.} 2023, \mnras, 519,
  4982, \dodoi{10.1093/mnras/stac3725}

\bibitem[{{Ding} {et~al.}(2024){Ding}, {Liao}, {Wu}, {Qi}, \&
  {Tang}}]{2024AA...691A..81D}
{Ding}, Y., {Liao}, S., {Wu}, Q., {Qi}, Z., \& {Tang}, Z. 2024, \aap, 691, A81,
  \dodoi{10.1051/0004-6361/202450967}

\bibitem[{{Doi} {et~al.}(2006){Doi}, {Fujisawa}, {Habe}, {Honma}, {Kawaguchi},
  {Kobayashi}, {Murata}, {Omodaka}, {Sudou}, \& {Takaba}}]{2006PASJ...58..777D}
{Doi}, A., {Fujisawa}, K., {Habe}, A., {et~al.} 2006, \pasj, 58, 777,
  \dodoi{10.1093/pasj/58.4.777}

\bibitem[{Euler(1776)}]{euler1776nova}
Euler, L. 1776, Novi commentarii academiae scientiarum Petropolitanae, 208

\bibitem[{{Fomalont} \& {Kopeikin}(2002)}]{2002evn..conf...53F}
{Fomalont}, E.~B., \& {Kopeikin}, S. 2002, in Proceedings of the 6th EVN
  Symposium, ed. E.~{Ros}, R.~W. {Porcas}, A.~P. {Lobanov}, \& J.~A. {Zensus},
  53

\bibitem[{{Fomalont} \& {Kopeikin}(2003)}]{2003ApJ...598..704F}
{Fomalont}, E.~B., \& {Kopeikin}, S.~M. 2003, \apj, 598, 704,
  \dodoi{10.1086/378785}

\bibitem[{{Gaia Collaboration} {et~al.}(2023){Gaia Collaboration}, {Vallenari},
  {Brown}, {Prusti}, {de Bruijne}, {Arenou}, {Babusiaux}, {Biermann},
  {Creevey}, {Ducourant}, {Evans}, {Eyer}, {Guerra}, {Hutton}, {Jordi},
  {Klioner}, {Lammers}, {Lindegren}, {Luri}, {Mignard}, {Panem}, {Pourbaix},
  {Randich}, {Sartoretti}, {Soubiran}, {Tanga}, {Walton}, {Bailer-Jones},
  {Bastian}, {Drimmel}, {Jansen}, {Katz}, {Lattanzi}, {van Leeuwen}, {Bakker},
  {Cacciari}, {Casta{\~n}eda}, {De Angeli}, {Fabricius}, {Fouesneau},
  {Fr{\'e}mat}, {Galluccio}, {Guerrier}, {Heiter}, {Masana}, {Messineo},
  {Mowlavi}, {Nicolas}, {Nienartowicz}, {Pailler}, {Panuzzo}, {Riclet}, {Roux},
  {Seabroke}, {Sordo}, {Th{\'e}venin}, {Gracia-Abril}, {Portell}, {Teyssier},
  {Altmann}, {Andrae}, {Audard}, {Bellas-Velidis}, {Benson}, {Berthier},
  {Blomme}, {Burgess}, {Busonero}, {Busso}, {C{\'a}novas}, {Carry}, {Cellino},
  {Cheek}, {Clementini}, {Damerdji}, {Davidson}, {de Teodoro}, {Nu{\~n}ez
  Campos}, {Delchambre}, {Dell'Oro}, {Esquej}, {Fern{\'a}ndez-Hern{\'a}ndez},
  {Fraile}, {Garabato}, {Garc{\'\i}a-Lario}, {Gosset}, {Haigron}, {Halbwachs},
  {Hambly}, {Harrison}, {Hern{\'a}ndez}, {Hestroffer}, {Hodgkin}, {Holl},
  {Jan{\ss}en}, {Jevardat de Fombelle}, {Jordan}, {Krone-Martins}, {Lanzafame},
  {L{\"o}ffler}, {Marchal}, {Marrese}, {Moitinho}, {Muinonen}, {Osborne},
  {Pancino}, {Pauwels}, {Recio-Blanco}, {Reyl{\'e}}, {Riello}, {Rimoldini},
  {Roegiers}, {Rybizki}, {Sarro}, {Siopis}, {Smith}, {Sozzetti}, {Utrilla},
  {van Leeuwen}, {Abbas}, {{\'A}brah{\'a}m}, {Abreu Aramburu}, {Aerts},
  {Aguado}, {Ajaj}, {Aldea-Montero}, {Altavilla}, {{\'A}lvarez}, {Alves},
  {Anders}, {Anderson}, {Anglada Varela}, {Antoja}, {Baines}, {Baker},
  {Balaguer-N{\'u}{\~n}ez}, {Balbinot}, {Balog}, {Barache}, {Barbato},
  {Barros}, {Barstow}, {Bartolom{\'e}}, {Bassilana}, {Bauchet}, {Becciani},
  {Bellazzini}, {Berihuete}, {Bernet}, {Bertone}, {Bianchi}, {Binnenfeld},
  {Blanco-Cuaresma}, {Blazere}, {Boch}, {Bombrun}, {Bossini}, {Bouquillon},
  {Bragaglia}, {Bramante}, {Breedt}, {Bressan}, {Brouillet}, {Brugaletta},
  {Bucciarelli}, {Burlacu}, {Butkevich}, {Buzzi}, {Caffau}, {Cancelliere},
  {Cantat-Gaudin}, {Carballo}, {Carlucci}, {Carnerero}, {Carrasco},
  {Casamiquela}, {Castellani}, {Castro-Ginard}, {Chaoul}, {Charlot}, {Chemin},
  {Chiaramida}, {Chiavassa}, {Chornay}, {Comoretto}, {Contursi}, {Cooper},
  {Cornez}, {Cowell}, {Crifo}, {Cropper}, {Crosta}, {Crowley}, {Dafonte},
  {Dapergolas}, {David}, {David}, {de Laverny}, {De Luise}, {De March}, {De
  Ridder}, {de Souza}, {de Torres}, {del Peloso}, {del Pozo}, {Delbo},
  {Delgado}, {Delisle}, {Demouchy}, {Dharmawardena}, {Di Matteo}, {Diakite},
  {Diener}, {Distefano}, {Dolding}, {Edvardsson}, {Enke}, {Fabre}, {Fabrizio},
  {Faigler}, {Fedorets}, {Fernique}, {Fienga}, {Figueras}, {Fournier},
  {Fouron}, {Fragkoudi}, {Gai}, {Garcia-Gutierrez}, {Garcia-Reinaldos},
  {Garc{\'\i}a-Torres}, {Garofalo}, {Gavel}, {Gavras}, {Gerlach}, {Geyer},
  {Giacobbe}, {Gilmore}, {Girona}, {Giuffrida}, {Gomel}, {Gomez},
  {Gonz{\'a}lez-N{\'u}{\~n}ez}, {Gonz{\'a}lez-Santamar{\'\i}a},
  {Gonz{\'a}lez-Vidal}, {Granvik}, {Guillout}, {Guiraud},
  {Guti{\'e}rrez-S{\'a}nchez}, {Guy}, {Hatzidimitriou}, {Hauser}, {Haywood},
  {Helmer}, {Helmi}, {Sarmiento}, {Hidalgo}, {Hilger}, {H{\l}adczuk}, {Hobbs},
  {Holland}, {Huckle}, {Jardine}, {Jasniewicz}, {Jean-Antoine Piccolo},
  {Jim{\'e}nez-Arranz}, {Jorissen}, {Juaristi Campillo}, {Julbe}, {Karbevska},
  {Kervella}, {Khanna}, {Kontizas}, {Kordopatis}, {Korn}, {K{\'o}sp{\'a}l},
  {Kostrzewa-Rutkowska}, {Kruszy{\'n}ska}, {Kun}, {Laizeau}, {Lambert},
  {Lanza}, {Lasne}, {Le Campion}, {Lebreton}, {Lebzelter}, {Leccia}, {Leclerc},
  {Lecoeur-Taibi}, {Liao}, {Licata}, {Lindstr{\o}m}, {Lister}, {Livanou},
  {Lobel}, {Lorca}, {Loup}, {Madrero Pardo}, {Magdaleno Romeo}, {Managau},
  {Mann}, {Manteiga}, {Marchant}, {Marconi}, {Marcos}, {Marcos Santos},
  {Mar{\'\i}n Pina}, {Marinoni}, {Marocco}, {Marshall}, {Martin Polo},
  {Mart{\'\i}n-Fleitas}, {Marton}, {Mary}, {Masip}, {Massari},
  {Mastrobuono-Battisti}, {Mazeh}, {McMillan}, {Messina}, {Michalik}, {Millar},
  {Mints}, {Molina}, {Molinaro}, {Moln{\'a}r}, {Monari}, {Mongui{\'o}},
  {Montegriffo}, {Montero}, {Mor}, {Mora}, {Morbidelli}, {Morel}, {Morris},
  {Muraveva}, {Murphy}, {Musella}, {Nagy}, {Noval}, {Oca{\~n}a}, {Ogden},
  {Ordenovic}, {Osinde}, {Pagani}, {Pagano}, {Palaversa}, {Palicio},
  {Pallas-Quintela}, {Panahi}, {Payne-Wardenaar}, {Pe{\~n}alosa Esteller},
  {Penttil{\"a}}, {Pichon}, {Piersimoni}, {Pineau}, {Plachy}, {Plum}, {Poggio},
  {Pr{\v{s}}a}, {Pulone}, {Racero}, {Ragaini}, {Rainer}, {Raiteri}, {Rambaux},
  {Ramos}, {Ramos-Lerate}, {Re Fiorentin}, {Regibo}, {Richards}, {Rios Diaz},
  {Ripepi}, {Riva}, {Rix}, {Rixon}, {Robichon}, {Robin}, {Robin}, {Roelens},
  {Rogues}, {Rohrbasser}, {Romero-G{\'o}mez}, {Rowell}, {Royer}, {Ruz Mieres},
  {Rybicki}, {Sadowski}, {S{\'a}ez N{\'u}{\~n}ez}, {Sagrist{\`a} Sell{\'e}s},
  {Sahlmann}, {Salguero}, {Samaras}, {Sanchez Gimenez}, {Sanna},
  {Santove{\~n}a}, {Sarasso}, {Schultheis}, {Sciacca}, {Segol}, {Segovia},
  {S{\'e}gransan}, {Semeux}, {Shahaf}, {Siddiqui}, {Siebert}, {Siltala},
  {Silvelo}, {Slezak}, {Slezak}, {Smart}, {Snaith}, {Solano}, {Solitro},
  {Souami}, {Souchay}, {Spagna}, {Spina}, {Spoto}, {Steele},
  {Steidelm{\"u}ller}, {Stephenson}, {S{\"u}veges}, {Surdej}, {Szabados},
  {Szegedi-Elek}, {Taris}, {Taylor}, {Teixeira}, {Tolomei}, {Tonello}, {Torra},
  {Torra}, {Torralba Elipe}, {Trabucchi}, {Tsounis}, {Turon}, {Ulla}, {Unger},
  {Vaillant}, {van Dillen}, {van Reeven}, {Vanel}, {Vecchiato}, {Viala},
  {Vicente}, {Voutsinas}, {Weiler}, {Wevers}, {Wyrzykowski}, {Yoldas}, {Yvard},
  {Zhao}, {Zorec}, {Zucker}, \& {Zwitter}}]{2023AA...674A...1G}
{Gaia Collaboration}, {Vallenari}, A., {Brown}, A.~G.~A., {et~al.} 2023, \aap,
  674, A1, \dodoi{10.1051/0004-6361/202243940}

\bibitem[{{Greisen}(2003)}]{2003ASSL..285..109G}
{Greisen}, E.~W. 2003, in Astrophysics and Space Science Library, Vol. 285,
  Information Handling in Astronomy - Historical Vistas, ed. A.~{Heck}, 109,
  \dodoi{10.1007/0-306-48080-8_7}

\bibitem[{Harris {et~al.}(2020)Harris, Millman, van~der Walt, Gommers,
  Virtanen, Cournapeau, Wieser, Taylor, Berg, Smith, Kern, Picus, Hoyer, van
  Kerkwijk, Brett, Haldane, del R{\'{i}}o, Wiebe, Peterson,
  G{\'{e}}rard-Marchant, Sheppard, Reddy, Weckesser, Abbasi, Gohlke, \&
  Oliphant}]{harris2020array}
Harris, C.~R., Millman, K.~J., van~der Walt, S.~J., {et~al.} 2020, Nature, 585,
  357, \dodoi{10.1038/s41586-020-2649-2}

\bibitem[{{Hunter}(2007)}]{2007CSE.....9...90H}
{Hunter}, J.~D. 2007, Computing in Science and Engineering, 9, 90,
  \dodoi{10.1109/MCSE.2007.55}

\bibitem[{{Hyland} {et~al.}(2022){Hyland}, {Reid}, {Ellingsen}, {Rioja},
  {Dodson}, {Orosz}, {Masson}, \& {McCallum}}]{2022ApJ...932...52H}
{Hyland}, L.~J., {Reid}, M.~J., {Ellingsen}, S.~P., {et~al.} 2022, \apj, 932,
  52, \dodoi{10.3847/1538-4357/ac6d5b}

\bibitem[{{Hyland} {et~al.}(2023){Hyland}, {Reid}, {Orosz}, {Ellingsen},
  {Weston}, {Kumar}, {Dodson}, {Rioja}, {Hankey}, {Yates-Jones}, {Natusch},
  {Gulyaev}, {Menten}, \& {Brunthaler}}]{2023ApJ...953...21H}
{Hyland}, L.~J., {Reid}, M.~J., {Orosz}, G., {et~al.} 2023, \apj, 953, 21,
  \dodoi{10.3847/1538-4357/acdbc5}

\bibitem[{{Kettenis} {et~al.}(2006){Kettenis}, {van Langevelde}, {Reynolds}, \&
  {Cotton}}]{2006ASPC..351..497K}
{Kettenis}, M., {van Langevelde}, H.~J., {Reynolds}, C., \& {Cotton}, B. 2006,
  in Astronomical Society of the Pacific Conference Series, Vol. 351,
  Astronomical Data Analysis Software and Systems XV, ed. C.~{Gabriel},
  C.~{Arviset}, D.~{Ponz}, \& S.~{Enrique}, 497

\bibitem[{{Lestrade} {et~al.}(1990){Lestrade}, {Rogers}, {Whitney}, {Niell},
  {Phillips}, \& {Preston}}]{1990AJ.....99.1663L}
{Lestrade}, J.~F., {Rogers}, A.~E.~E., {Whitney}, A.~R., {et~al.} 1990, \aj,
  99, 1663, \dodoi{10.1086/115447}

\bibitem[{{Petrov} \& {Kovalev}(2025)}]{2025ApJS..276...38P}
{Petrov}, L.~Y., \& {Kovalev}, Y.~Y. 2025, \apjs, 276, 38,
  \dodoi{10.3847/1538-4365/ad8c36}

\bibitem[{{Reid}(2022)}]{2022PASP..134l3001R}
{Reid}, M.~J. 2022, \pasp, 134, 123001, \dodoi{10.1088/1538-3873/acabe6}

\bibitem[{{Reid} {et~al.}(2009){Reid}, {Menten}, {Brunthaler}, {Zheng},
  {Moscadelli}, \& {Xu}}]{2009ApJ...693..397R}
{Reid}, M.~J., {Menten}, K.~M., {Brunthaler}, A., {et~al.} 2009, \apj, 693,
  397, \dodoi{10.1088/0004-637X/693/1/397}

\bibitem[{{Rioja} \& {Dodson}(2020)}]{2020AARv..28....6R}
{Rioja}, M.~J., \& {Dodson}, R. 2020, \aapr, 28, 6,
  \dodoi{10.1007/s00159-020-00126-z}

\bibitem[{{Rioja} {et~al.}(2017){Rioja}, {Dodson}, {Orosz}, {Imai}, \&
  {Frey}}]{2017AJ....153..105R}
{Rioja}, M.~J., {Dodson}, R., {Orosz}, G., {Imai}, H., \& {Frey}, S. 2017, \aj,
  153, 105, \dodoi{10.3847/1538-3881/153/3/105}

\bibitem[{{Rioja} {et~al.}(2002){Rioja}, {Porcas}, {Desmurs}, {Alef},
  {Gurvits}, \& {Schilizzi}}]{2002evn..conf...57R}
{Rioja}, M.~J., {Porcas}, R.~W., {Desmurs}, J.~F., {et~al.} 2002, in
  Proceedings of the 6th EVN Symposium, ed. E.~{Ros}, R.~W. {Porcas}, A.~P.
  {Lobanov}, \& J.~A. {Zensus}, 57, \dodoi{10.48550/arXiv.astro-ph/0207210}

\bibitem[{{Rioja} {et~al.}(1997){Rioja}, {Stevens}, {Gurvits}, {Alef},
  {Schilizzi}, {Sasao}, \& {Asaki}}]{1997VA.....41..213R}
{Rioja}, M.~J., {Stevens}, E., {Gurvits}, L., {et~al.} 1997, Vistas in
  Astronomy, 41, 213, \dodoi{10.1016/S0083-6656(97)00008-1}

\bibitem[{Rodrigues(1840)}]{JMPA_1840_1_5__380_0}
Rodrigues, O. 1840, Journal de Math\'ematiques Pures et Appliqu\'ees, 1e
  s{\'e}rie, 5, 380.
\newblock \url{http://www.numdam.org/item/JMPA_1840_1_5__380_0/}

\bibitem[{{The pandas development Team}(2024)}]{2022zndo...3509134T}
{The pandas development Team}. 2024, {pandas-dev/pandas: Pandas}, v2.2.3,
  Zenodo, \dodoi{10.5281/zenodo.3509134}

\bibitem[{{Virtanen} {et~al.}(2020){Virtanen}, {Gommers}, {Oliphant},
  {Haberland}, {Reddy}, {Cournapeau}, {Burovski}, {Peterson}, {Weckesser},
  {Bright}, {van der Walt}, {Brett}, {Wilson}, {Millman}, {Mayorov}, {Nelson},
  {Jones}, {Kern}, {Larson}, {Carey}, {Polat}, {Feng}, {Moore}, {VanderPlas},
  {Laxalde}, {Perktold}, {Cimrman}, {Henriksen}, {Quintero}, {Harris},
  {Archibald}, {Ribeiro}, {Pedregosa}, {van Mulbregt}, \& {SciPy 1. 0
  Contributors}}]{2020NatMe..17..261V}
{Virtanen}, P., {Gommers}, R., {Oliphant}, T.~E., {et~al.} 2020, Nature
  Methods, 17, 261, \dodoi{10.1038/s41592-019-0686-2}

\bibitem[{{Wenger} {et~al.}(2000){Wenger}, {Ochsenbein}, {Egret}, {Dubois},
  {Bonnarel}, {Borde}, {Genova}, {Jasniewicz}, {Lalo{\"e}}, {Lesteven}, \&
  {Monier}}]{simbad}
{Wenger}, M., {Ochsenbein}, F., {Egret}, D., {et~al.} 2000, \aaps, 143, 9,
  \dodoi{10.1051/aas:2000332}

\bibitem[{Zhang(2025)}]{jingdong_zhang_2025_15030432}
Zhang, J. 2025, FrdCHK/serial-MultiView: First release, v1.0.0,  Zenodo,
  \dodoi{10.5281/zenodo.15030432}

\bibitem[{{Zhang} {et~al.}(2024){Zhang}, {Zhang}, {Xu}, {Liu}, {Chen}, {Ding},
  {Jiang}, {Sun}, {Wang}, {Cui}, {Wen}, {Mai}, {Li}, {Shu}, \&
  {Huang}}]{2024MNRAS.529.2062Z}
{Zhang}, J., {Zhang}, B., {Xu}, S., {et~al.} 2024, \mnras, 529, 2062,
  \dodoi{10.1093/mnras/stae705}

\end{thebibliography}
\bibliographystyle{aasjournal}



\end{document}